\documentclass[oneside]{elsart}
\usepackage{graphicx}
\usepackage{epstopdf}
\usepackage[figuresright]{rotating}
\usepackage{amsmath}

\def\intablecenterline#1{\hfill#1\hfill} 
\def\und{} 
\def\ETTunderline#1{\ifmmode\let\und=\underline\else 
\let\und=\underbar\fi\und{#1}}

\def\mathline{\ifmmode \vert\else$\vert$\fi}

\def\ETTstretch#1{\message{[[Sorry, you can't use Stretch with ETT]]}} 
 
\def\ETTexpand#1{\message{[[Sorry, you can't use Expand with ETT]]}} 
 
\def\xbackslash{\ifmmode \backslash\else$\backslash$\fi}

\def\quoteswitch{} 
\def\leftquotemarks{\ifmmode{}{}''\else``%
\let\quoteswitch=\rightquotemarks\fi} 
\def\rightquotemarks{\ifmmode{}{}''\else"\let\quoteswitch=\leftquotemarks\fi} 
\let\quoteswitch=\leftquotemarks 
\def\p$&-\${\overline{p}} 
\def\ETTstack#1#2{\ifmmode\ifx\int#1\displaystyle#1\limits#2\else 
\ifx\xint#1\displaystyle\int\limits#2\else 
\ifx\sum#1\displaystyle#1\limits#2\else%
\ifx\xsum#1\displaystyle\sum\limits#2\else 
\mathop{#1}\limits#2\fi\fi\fi\fi\else#1\mathop{#2}\limits\fi} 
 
\def\ETThalign#1{{\let\centerline=\intablecenterline 
\ifmmode\vcenter{\everycr={\noalign{\vskip1.5pt}}%
\halign{$\strut##$&$\strut##$&$\strut##$&$\strut##$&$\strut##$&%
$\strut##$&$\strut##$&$\strut##$&$\strut##$&$\strut##$&$\strut##$&%
$\strut##$&$\strut##$&$\strut##$&$\strut##$&$\strut##$&$\strut##$&%
$\strut##$&$\strut##$&$\strut##$\cr#1\crcr}\everycr={}}\else%
\hbox{$\vcenter{\everycr={\noalign{\vskip1.5pt}} 
\halign{\strut##&\strut##&\strut##&\strut##&%
\strut##&\strut##&\strut##&\strut##&\strut##&\strut##&\strut##&%
\strut##&\strut##&\strut##&\strut##&\strut##&\strut##&\strut##&%
\strut##&\strut##\cr#1\crcr}\everycr={}}$}\fi}} 
 
\gdef\ETTcolumn#1{\ifmmode%
\hbox{$\vcenter{\everycr{\noalign{\vskip-2pt}}%
\let\centerline=\intablecenterline 
\halign{\strut\hfill$##$\hfill\cr 
#1\crcr}\vskip-2pt}$}\else 
\hbox{$\vcenter{\offinterlineskip%
\let\centerline=\intablecenterline 
\halign{\strut\hfill##\hfill\cr 
#1\crcr}}$}\fi} 
 
\def\ETTtable#1{\ifmmode\vcenter{\everycr={\noalign{\vskip1.5pt}}%
\let\centerline=\intablecenterline%
\halign{\hfil$\vcenter{\hbox{$\strut##$}}$\hfil\ &\ %
\hfil$\vcenter{\hbox{$\strut##$}}$\hfil\ &\ %
\hfil$\vcenter{\hbox{$\strut##$}}$\hfil\ &\ %
\hfil$\vcenter{\hbox{$\strut##$}}$\hfil\ &\ %
\hfil$\vcenter{\hbox{$\strut##$}}$\hfil\ &\ %
\hfil$\vcenter{\hbox{$\strut##$}}$\hfil\ &\ %
\hfil$\vcenter{\hbox{$\strut##$}}$\hfil\ &\ %
\hfil$\vcenter{\hbox{$\strut##$}}$\hfil\ &\ %
\hfil$\vcenter{\hbox{$\strut##$}}$\hfil\ \cr 
#1\crcr}\everycr={}}\vrule depth3pt width0pt\relax\else 
\hbox{$\vcenter{\everycr={\noalign{\vskip1.5pt}}%
\let\centerline=\intablecenterline%
\halign{\hfil$\vcenter{\hbox{\strut##}}$\hfil\ &\ %
\hfil$\vcenter{\hbox{\strut##}}$\hfil\ &\ %
\hfil$\vcenter{\hbox{\strut##}}$\hfil\ &\ %
\hfil$\vcenter{\hbox{\strut##}}$\hfil\ &\ %
\hfil$\vcenter{\hbox{\strut##}}$\hfil\ &\ %
\hfil$\vcenter{\hbox{\strut##}}$\hfil\ &\ %
\hfil$\vcenter{\hbox{\strut##}}$\hfil\ &\ %
\hfil$\vcenter{\hbox{\strut##}}$\hfil\ &\ %
\hfil$\vcenter{\hbox{\strut##}}$\hfil\ \cr 
#1\crcr}\everycr={}}$}\vrule depth3pt width0pt\relax\fi} 
 
\def\ETTast{{\raise.25ex\hbox{$\ast$}}} 
 
\newif\ifinmath 
\newbox\superbox 
\newbox\superboxtwo 
 
\def\changetomath{$} 
 
\def\ETTsuperimpose#1#2{\ifmmode\let\check=\changetomath%
\else\let\check=\relax\fi%
\setbox\superbox=\hbox{\check#1\check}%
\setbox\superboxtwo=\hbox{\check#2\check}%
\ifdim\wd\superbox>\wd\superboxtwo%
\copy\superbox\hskip-\wd\superbox\hbox to 
\wd\superbox{\hfill\check#2\check\hfill}%
\else%
\copy\superboxtwo\hskip-\wd\superboxtwo\hbox to 
\wd\superboxtwo{\hfill\check#1\check\hfill}\fi} 
 
\def\hb{{\hskip3pt}} 
\let~=\hb

\newcount\moveup \newcount\moveover 
 
\def\lookagain{\futurelet\next\parser} 
 
\def\parser#1{\def\endb{}\ifx\next /\let\go=\relax \else\ifx\next r 
\global\advance\moveover by22\else 
\ifx\next l\global\advance\moveover by-22 
\else \ifx\next u\global\advance\moveup by22 
\else \ifx\next d\global\advance\moveup by-22\fi\fi\fi\fi 
\let\go=\lookagain\fi\go} 
 
\def\ETTadjust#1#2{\moveover=0 \moveup=0\setbox0=\hbox{\lookagain#1/}%
\divide\moveover by10 \divide\moveup by10\setbox1=\hbox{#2}%
\vtop to0pt{\hbox to\wd1{\hskip\moveover pt\raise\moveup pt\hbox{#2}\hss}%
\vss}} 
 
\newcount\upcounter 
\newcount\overcounter 
 
\newif\ifaz \newif\ifrz \newif\iflz \newif\ifdz \newif\ifuz 
 
\def\nonglobalreset{\azfalse\lzfalse\uzfalse%
\dzfalse\rzfalse}%
\long\gdef\ETTbox #1 #2{\def\parse{#1}%
\nonglobalreset\expandafter\looker\parse{}{}{}\boxer{\vbox{%
\let\centerline=\intablecenterline\halign{%
\strut##\cr#2\crcr}}}} 
 
\long\def\boxer#1{\ifaz\rztrue\lztrue\uztrue\dztrue\fi%
\hbox{$\,\,\vcenter{\ifuz\hrule\fi\hbox{\iflz\vrule\fi%
\hskip2pt\vbox{\vskip2pt\vbox{#1}\vskip2pt}%
\hskip2pt\ifrz\vrule\fi}\ifdz\hrule\fi}\,\,$}} 
 
\def\looker #1#2#3#4{%
\ifx#1A\aztrue\else\ifx#2a\aztrue\else
\ifx#1R\rztrue\else 
\ifx#2R\rztrue\else 
\ifx#3R\rztrue\else 
\ifx#4R\rztrue\fi\fi\fi\fi\fi 
\ifx#1L\lztrue\else 
\ifx#2L\lztrue\else 
\ifx#3L\lztrue\else 
\ifx#4L\lztrue\fi\fi\fi\fi 
\ifx#1U\uztrue\else 
\ifx#2U\uztrue\else 
\ifx#3U\uztrue\else 
\ifx#4U\uztrue\fi\fi\fi\fi 
\ifx#1D\dztrue\else 
\ifx#2D\dztrue\else 
\ifx#3D\dztrue\else 
\ifx#4D\dztrue\fi\fi\fi\fi\fi} 
 
\def\tie{\ifmmode\sim\else 
\hbox{\vtop to0pt{\hbox{\lower5pt\hbox{\~{}}}\vss}}\fi} 
\def\<{\ifmmode <\else $<$\fi} 
\def\>{\ifmmode >\else $>$\fi} 
 
\def\caret{\ifmmode{\vtop to0pt{\hbox{\lower7pt\hbox{$\hat{\vphantom{.}}$}} 
\vss}}\else \^{}\fi} 
 
\def\lcurlybracket{\ifmmode\{\else $\{$\fi} 
\def\rcurlybracket{\ifmmode\}\else $\}$\fi} 
 
\newdimen\tempdimen 
 
\def\smdarkP{\ \hbox{\global\setbox0=\hbox{\smallbackp}\vrule height\ht0 
depth\dp0 width0pt\vrule\hskip.6pt\vrule}\tempdimen=\ht0%
\advance\tempdimen by-.4pt%
\hskip-1pt\raise1.4pt\hbox{$\scriptstyle\bullet$}%
\hskip-2pt\llap{\raise\tempdimen\hbox to3pt{\hrulefill}}\ } 
 
\def\xxdarkP{\ \hbox{\global\setbox0=\hbox{\P}\vrule height\ht0 
depth\dp0 width0pt\vrule\hskip.6pt\vrule}\tempdimen=\ht0%
\advance\tempdimen by-.4pt\hskip-1pt\raise3pt\hbox{$\displaystyle\bullet$}%
\hskip-2pt\llap{\raise\tempdimen\hbox to4pt{\hrulefill}}\ }

\def\smallbackp{{\smallsymbol\char'173}}

\def\return{\hbox{\ \unskip{$\bf\leftarrow$\hskip-.5pt\raise2.4pt%
\hbox{\vrule height 2.5pt}}}}

\def\xxTM{\raise1ex\hbox{\amrseven TM}} 
 
\def\TMfive{\raise1ex\hbox{\amrfive TM}}

\def\xxeqcirc{\buildrel \lower1.5pt\hbox{$\scriptstyle\circ$}\over =} 
\def\smeqcirc{\buildrel \lower1.5pt\hbox{$\scriptscriptstyle\circ$}\over{\scriptstyle =}}

\def\xxeqdkcirc{\buildrel \lower1.5pt\hbox{$\scriptscriptstyle\bullet$}\over =} 
\def\smeqdkcirc{\buildrel \lower1.5pt\hbox{$\scriptscriptstyle\bullet$}\over{\scriptstyle =}}

\def\xxtridots{\ \unskip\raise4pt\hbox 
to1em{\hfil.\hfil}\llap{\hbox to1em{\hfil.\ .\hfil}}} 
 
\def\smtridots{\ \unskip\raise3pt\hbox 
to1em{\hfil.\hfil}\llap{\hbox to1em{\hfil.$\,$.\hfil}}}

\def\sqr#1#2{{\vcenter{\hrule height.#2pt 
\hbox{\vrule width.#2pt height#1pt\kern#1pt 
\vrule width.#2pt} 
\hrule height.2pt}}}

\def\xint{{\mathchoice{\int}{\displaystyle\int}{\int}{\int}}} 
\def\xsum{{\mathchoice{\sum}{\displaystyle\sum}{\sum}{\sum}}} 
 
\def\GermanS{\ifmmode\hbox{\ss}\else\ss\fi} 
\def\primeaccent{\ifmmode\hbox{\rm\char19}\else{\rm\char19}\fi} 
\def\underaccent{\ifmmode\hbox{\rm\char24}\else{\rm\char24}\fi} 
\def\EnglishPound{\ifmmode\hbox{\it\$}\,\else{\it\$}\fi} 
 
\newcommand{\be}{\begin{equation}}
\newcommand{\ee}{\end{equation}}
\newcommand{\bea}{\begin{eqnarray}}
\newcommand{\eea}{\end{eqnarray}}
\newcommand{\beas}{\begin{eqnarray*}}
\newcommand{\eeas}{\end{eqnarray*}}
\newcommand{\bi}{\begin{itemize}}
\newcommand{\ei}{\end{itemize}}
\newcommand{\bn}{\begin{enumerate}}
\newcommand{\en}{\end{enumerate}}

\def\lambar{{ \lambda \mkern-10mu\raise.5ex\hbox{--} }}

\def\thus{{ .. \mkern-7.5mu\raise.9ex\hbox{.} }\  }

\def\ba2#1#2{${\overline{#1}}^{#2}$}
\def\anti#1#2{\vbox{\ialign{##\crcr
     \hrulefill$\smash{\phantom{\scriptstyle#2}}$\crcr
     \noalign{\kern-1pt\nointerlineskip\vskip 0.25ex}
     $\hfil{#1}^{#2}\hfil$\crcr}}}
\def\anth#1#2{\vbox{\ialign{##\crcr
    \hrulefill$\smash{\phantom{\scriptstyle#2}}$\crcr
    \noalign{\kern-0.5pt\nointerlineskip\vskip 0.25ex}
    $\hfil{#1}^{#2}\hfil$\crcr}}}
\def\rless{{ r \mkern-.5mu\raise-.2ex\hbox{\tiny{<}} }}
\def\rmore{{ r \mkern-.5mu\raise-.2ex\hbox{\tiny{>}} }}

\newcommand{\NOT}{{\rm\bf\goodfontB NOT }}

\newcommand{\QDENS}{{\rm\bf\goodfontB QDENSITY\,}}
\newcommand{\QCMPI}{{\rm\bf\goodfontB QCMPI\,}}

\def\goodfontB{\usefont{T1}{phv}{n}{n}\fontsize{9pt}{9pt}\selectfont}

\newcommand{\oracle}{{\cal \bf O\rm racle}}

\usepackage[matrix,frame,arrow]{xy}
\usepackage{amsmath}

\begin{document}
\begin{frontmatter}
\title{QCMPI: A PARALLEL ENVIRONMENT FOR QUANTUM COMPUTING }

\author[pit]{Frank Tabakin}
\and
\author[spain]{Bruno Juli\'a-D\'{\i}az}

\address[pit]{Department of Physics and Astronomy \\
University of Pittsburgh\\
Pittsburgh,  PA, 15260}

\address[spain]{Departament de Estructura i Constituents de la Materia\\
Universitat de Barcelona\\
Diagonal 647\\
08028 Barcelona (Spain)
}

\begin{abstract}
\QCMPI \  is a quantum computer (QC) simulation package written in 
Fortran 90 with parallel processing capabilities. It is an accessible 
research tool that permits rapid evaluation of quantum algorithms for 
a large number of qubits and for various  ``noise" scenarios. 
The prime motivation for developing \QCMPI\ is to facilitate numerical 
examination of not only how QC algorithms work, but also to include noise,
decoherence, and attenuation effects and to evaluate the efficacy of 
error correction schemes. The present work builds on an earlier Mathematica 
code \QDENS, which is mainly a pedagogic tool. In that earlier work, 
although the density matrix formulation was featured, the description 
using state vectors was also provided.  In  \QCMPI,  the stress is on 
state vectors, in order to employ a large number of qubits. The parallel 
processing feature is implemented by using the Message-Passing Interface 
(MPI) protocol. A description of how to spread the wave function components 
over many processors is provided, along with how to efficiently describe 
the action of general one-  and  two-qubit operators on these state vectors. 
These operators include the standard Pauli, Hadamard, CNOT and CPHASE gates 
and also Quantum Fourier transformation. These operators make up the actions 
needed in QC. Codes for Grover's search and Shor's factoring algorithms 
are provided as examples. A major feature of this work is that concurrent 
versions of the algorithms can be evaluated with each version subject to 
alternate noise effects, which corresponds to the idea of solving a 
stochastic Schr\"{o}dinger equation. The density matrix for the ensemble of 
such noise cases is constructed using parallel distribution methods to
evaluate its eigenvalues and  associated entropy. Potential 
applications of this powerful tool include studies of the stability and
correction of QC processes using Hamiltonian based dynamics.

\vspace{1pc}
\end{abstract}
\end{frontmatter}

\noindent{\bf Program Summary}

{\it Title of program:} QCMPI\\
{\it Catalogue identifier:}\\
{\it Program summary URL:} http://cpc.cs.qub.ac.uk/summaries\\
{\it Program available from:} CPC Program Library, Queen's University of Belfast, N. Ireland\\
{\it Operating systems:} Any system that supports Fortran 90 and MPI;
developed and tested at the Pittsburgh Supercomputer Center, at the Barcelona 
Supercomputer (BSC/CNS) and on multi-processor Macs and PCs. For cases where distributed density
matrix evaluation is invoked,  the BLACS and SCALAPACK packages are needed, \\
{\it Programming language used:} Fortran 90 and MPI\\
{\it Number of bytes in distributed program, including test code and 
documentation:  }\\
{\it Distribution format:} tar.gz\\
{\it Nature of Problem:} Analysis  of quantum computation algorithms and the effects of noise. \\
{\it Method of Solution:} A Fortran 90/MPI  package is provided that contains
modular commands to create and analyze quantum circuits. Shor's factorization
and Grover's search algorithms are explained in detail. Procedures for
distributing state vector amplitudes over processors and for solving
concurrent (multiverse) cases with noise effects are implemented.  Density 
matrix and entropy evaluations are provided in both single and parallel versions. \\

\newpage
\setlength{\parskip}{0.1cm}
\tableofcontents
\setlength{\parskip}{\baselineskip}
\newpage
\section{INTRODUCTION}

Achieving a realistic Quantum Computer (QC)~\cite{Nielsen,Preskill}  requires 
the control, measurement, and stability of simple quantum systems called qubits.
A qubit is any system with two accessible states which can form a quantum
ensemble. That ensemble can be manipulated to store and process information.  
Since quantum states can exist as superpositions of many possibilities, and 
since an isolated quantum system propagates without loss of quantum phases, 
a QC provides the advantage of being a ``massively parallel" device and having 
enhanced probability for solving difficult, otherwise intractable,
problems. That enhancement is generated by constructive quantum interference. 
This ideal situation can be disrupted by external effects, which can cause the 
quantum system to loose its quantum interference capabilities--this is called 
decoherence and loss of entanglement. In addition, uncontrolled random pulses 
(noise~\footnote{There are various types of noise, such as thermal noise. We 
use the term noise in a generic sense, although specific noise models can be 
incorporated into \QCMPI. }) could strike the QC during its controlled
performance and thereby its operations or gates can be less than perfect. 
  
To gauge the efficacy of a QC, even when influenced by such external
environmental effects, and to evaluate the positive influence of error 
correction~\cite{QEC} steps, it is important to have large scale QC simulations.  Such 
simulations can only represent a small part of the full  ``massively parallel"
quantum ensemble dynamics, since a real QC goes way beyond the capabilities of 
any classical computer.  Nevertheless,  it seems natural to invoke the best, 
most parallel and largest memory computers we have available. Therefore, we 
embarked on developing a Fortran 90 parallel computer QC simulation, starting 
with the basic QC algorithms of quantum searching~\cite{Grover} and factorization~\cite{Shor}.
Other authors have also attacked this problem to good
effect~\cite{prevQC0,prevQC1,prevQC2,prevQC3,prevQC4}. Nevertheless, there is a need for a 
generally available, well-documented, and easy to use supercomputer version, 
to encourage others to contribute their own advances.  In addition, we have 
developed a broader range of applications~\footnote{Teleportation and superdense 
coding programs are also available, but were omitted for brevity.} and
supercomputer techniques than previously available.  An important feature 
of our work is that we invoke the algorithms on concurrent groups of 
processors, which are then subject to different noise. Then, the overall 
density matrix is constructed as an ensemble average over these noise groups. 
The density matrix can be stored on a grid of processors and its eigenvalues
found using parallel codes,  thereby avoiding the pitfalls
of overly large matrix storage. Thus, we can evaluate the entropy, and indeed 
sub-entropies, for the dynamic evolution of a QC process 
in a simulation of a real world environment.

Our code is called \QCMPI \  to indicate that it is a QC simulation based 
on the Message-Passing Interface (MPI)~\cite{MPI,openmpi}. It is a Fortran 90
simulation of a Quantum Computer that is both flexible and an
improvement over earlier such works~\cite{prevQC0,prevQC1,prevQC2,prevQC3,prevQC4}.
The flexibility is generated by a modular approach to all of the
initializations, operators, gates, and measurements, which then
can be readily used to describe the basic QC
Teleportation~\cite{Teleportation}, Superdense coding~\cite{superdense}, 
Grover's search~\cite{Grover} and Shor's factoring~\cite{Shor} algorithms. 
We also adopt a state vector,\!~\footnote{A state vector requires arrays of 
size  $2^{n_q},$ whereas a density matrix has a much larger size $2^{n_q}
\times 2^{n_q}.$ Here $n_q$ denotes the number of qubits.}\! rather than a 
density matrix~\cite{QDENSITY},  approach to facilitate representing a large 
number of qubits in a manner that allows for general treatments, such as 
handling the dynamics stipulated by realistic Hamiltonians. We include 
environmental effects by introducing  random stochastic interactions in 
separate groups of processors, that we dub multiverses.

In section~\ref{sec2}, we introduce qubit state vectors along with various 
state vector notations. We stress that a wave function component description 
allows for changes induced by simple one-body operators such as local quantum 
gates and also one-body parts of Hamiltonians. Examples are provided in 
section~\ref{sec3} of the affect of a general one-body operator on both  two 
and more qubit systems. Expansion in a computational basis, using equivalent 
decimal and binary labels, is used to demonstrate the role of operators on the 
state vector amplitudes. It is shown how to distribute a wave function over 
numerous processors and how to handle the fact that a one-body operator acts 
on wave function amplitudes in an manner that not only modifies amplitudes 
stored  on a given processor, but also affects amplitudes seated on other 
processors. Criteria for locating the processors involved in these classes 
of operators are derived. Understanding this combination of effects; namely, 
wave function distribution and the alteration of that distribution due to the 
action of a one-body operator, is central to all subsequent developments. It 
is handled by careful MPI invocations and serves as a model for the extension 
to multi-qubit operations.

In section~\ref{sec4}, the MPI manipulations described earlier for the one
qubit case are generalized and then the layout for the two-qubit  operator 
alterations of  the quantum amplitudes are clarified. With that result in
hand, the particular two-qubit gates {\bf CNOT} and {\bf CPHASE} are readily constructed, 
as are two-body Hamiltonians for dynamical applications. Generalization to 
three-qubit operators, in particular to the Toffolli gate, are obvious.

In  section~\ref{sec5}, Grover's algorithm is discussed and it is shown how
\QCMPI allows one to simulate up to 30 qubits, (depending on the number of 
processors and available memory) in a reasonable time.  

Shor's algorithm is simulated using \QCMPI\  as discussed in
section~\ref{sec5}. Several standard codes that handle large-number modular 
and continued fraction manipulations are provided, but the heart of this case
is the Quantum Fourier Transform (QFT) and an associated projective measurement.
The QFT is generated by a chain of Hadamards and CNOT gates acting on a
multi-qubit register. It is shown how to do a QFT with wave function
components distributed over many processors. Here the benefit of using MPI is 
dramatic.

In section~\ref{sec7},  the procedures invoked to describe parallel universes, 
subject to stochastic noise, is explained for both the Grover and Shor
algorithms. For brevity, similar application to teleportation and superdense 
coding are omitted here (although also implemented using \QCMPI).
Also in  section~\ref{sec7},  the construction and evaluation of a density 
matrix is discussed in two ways. In one way, the full density matrix is 
stored on the master processor and its eigenvalues and the associated entropy 
is evaluated using a linear code subroutine.  In the second, more general way,
the density matrix is spread over many processors on a BLACS constructed 
processor grid and eigenvalues and entropy determined using the parallel 
library SCALAPACK~\cite{scala}. The later version reduces the storage needs and enhances speed.

A brief description of the included routines is given in section~\ref{sec8}, 
and finally some conclusions and future developments 
are discussed in section~\ref{sec9}.

 
\section{STATES  }
\label{sec2}
\subsection{One-Qubit States  }

The state of a quantum system is described by a wave function which
in general depends on the space or momentum coordinates of the
particles and on time. In Dirac's representation-independent
notation, the state of a system is a vector in an abstract Hilbert
space $\mid \Psi(t) \rangle$, which depends on time, but in that form one
makes no choice between the coordinate or momentum space
representation. The transformation between the space and momentum
representation is contained in a transformation bracket. The two
representations are thus related by Fourier transformation, which is the
way Quantum Mechanics builds localized wave packets. In this manner,
uncertainty principle limitations on our ability to measure
coordinates and momenta simultaneously with arbitrary precision are
embedded into Quantum Mechanics (QM). This fact leads to operators,
commutators, expectation values and, in the special cases when a
physical attribute can be precisely determined, eigenvalue equations
with Hermitian operators. That is the content of many quantum texts.
Our purpose is now to see how to define a state vector, to
describe systems or ensembles of qubits as needed for quantum
computing. Thus, the degrees of freedom associated with change in 
location are suppressed and the focus is on the two-state aspect.

Spin, which is the most basic example of two-valued quantum attribute, is
missing from a spatial description. This subtle degree of freedom,
whose existence is deduced, inter alia, by analysis of the Stern-Gerlach
experiment, is an additional Hilbert space vector feature. For
example, for a single spin 1/2 system the wave function including
both space and spin aspects is: \be \Psi(\vec{r}_1,  t) \mid s\
m_s\rangle, \ee where $\mid s \ m_s\rangle$ denotes a state that is
simultaneously an eigenstate of the particle's total spin operator
$s^2 = s_x^2 +s_y^2+s_z^2$, and of its spin component operator
$s_z$. That is \be s^2 \mid s m_s\rangle = \hbar^2 s (s+1) \mid s m_s\rangle
\qquad s_z \mid s m_s\rangle = \hbar m_s \mid s m_s\rangle \,. \ee For a spin
1/2 system, we denote the spin up state as $\mid s m_s\rangle\rightarrow
\mid \frac{1}{2},\frac{1}{2}\rangle \equiv \mid 0\rangle$, and the spin down
state as $\mid s m_s\rangle\rightarrow \mid \frac{1}{2},-\frac{1}{2}\rangle
\equiv \mid 1\rangle.$ 

 A simpler, equivalent representation is as a two component amplitude 
\be  
\mid 0\rangle\rightarrow \left(
\begin{array}{l}
 1 \\
 0
\end{array}
\right)  \, ,
\qquad
\mid 1 \rangle\rightarrow{\bf \left(
\begin{array}{l}
 0 \\ 1
\end{array}
\right)} \, . 
\label{spinor1} 
\ee
    
This matrix representation can be used to describe the two states of
any quantum system  and is not restricted to the spin attribute. In 
this matrix representation, the Pauli matrices $ \vec{\sigma}$ are: 
\footnote{ $\vec{s} =\frac{\hbar}{2} \vec{\sigma}$ }
 
\be
 \sigma_z  \longrightarrow
 \left(
\begin{array}{lc}
1 & \ \ 0\\
0 & -1
\end{array}\right)  \, ,
\qquad
  \sigma_x   \longrightarrow
\left(
\begin{array}{lc}
0 & \ \ 1\\
1 &  0
\end{array}\right) \, ,
\qquad
  \sigma_y  \longrightarrow \left(
\begin{array}{lc}
0 & -i \\
i & \ \ 0
\end{array}\right) \, .
 \label{Pauli}
\ee  
These are all Hermitian matrices $\sigma_i =\sigma^\dagger_i$.  Along with the 
unit operator ${\cal I}\equiv \sigma_0$
\be
{\cal I}\equiv  \sigma_0  \longrightarrow
  \left(
\begin{array}{lc}
1 & \ \ 0\\
0 &  \ \ 1
\end{array}\right)  \, ,
 \label{unitM} 
\ee  
any operator acting on a qubit can be expressed as a combination of  Pauli operators.
 
Operators on multi-qubit states can be expressed as linear combinations of the
tensor product~\footnote{  A tensor product of two matrices $A\otimes B$  is 
defined by the rule: 
$\langle q_i , q_j \mid~ A\otimes B~ \mid~q_s, q_t\rangle \equiv  \langle q_i  \mid A \mid q_s  \rangle\langle  q_j \mid B \mid  q_t\rangle,$
 with obvious generalization to multi-qubit operators.} of the Pauli
operators. For example, a general operator $\Omega$ can be expressed as
\be
\Omega = \sum_{ i_1=0 }^{3} \cdots  \sum_{ i_{n_q}=0 }^{3}\     
\beta_{ i_1, i_2 \cdots  i_{n_q} }  
\   [ \sigma_{i_1}  \otimes \sigma_{i_2}  
\cdots   \otimes \sigma_{i_{n_q}} ],
\ee  
where $ \beta_{ i_1, i_2 \cdots  i_{n_q} } $ is in general a set of 
complex numbers, but are real numbers for hermitian $\Omega.$

A one qubit state is a superposition of the two states associated 
with the above $0$ and $1$ bits: 
\be 
\mid \Psi\rangle =C_0  \mid 0\rangle+C_ 1  \mid 1\rangle, 
\ee 
where $C_0 \equiv \langle0\mid \Psi\rangle$ and 
$C_1 \equiv \langle1\mid \Psi\rangle$ are complex probability amplitudes 
for finding the particle with spin up or spin down, respectively. The 
normalization of the state $\langle\Psi\mid\Psi\rangle =1$, yields 
$\mid C_0 \mid^2 + \mid C_1 \mid^2=1$. Note that the spatial aspects of 
the wave function are being suppressed; which corresponds to the particles 
being in a fixed location, such as at quantum dots~\footnote{When these 
separated systems interact, one might need to restore the spatial aspects 
of the full wave function.}. A $2 \times 1$ matrix representation of this 
one-qubit state is thus:
\be
\mid \Psi \rangle \rightarrow  \left(
\begin{array}{l}
 C_0 \\
 C_1
\end{array}
\right)    \, .   
\label{amp1} 
\ee 

An essential point is that a QM system can exist in a
superposition of these two bits; hence, the state is called a
quantum-bit or ``qubit." Although our discussion uses the notation
of a system with spin 1/2, it should be noted again that the same discussion
applies to any two distinct states that can be associated with
$\mid 0\rangle $ and $ \mid 1\rangle$.  

\subsection{ Multi-qubit States}

A quantum computer involves more than one qubit; therefore, we generalize the 
previous section to multi-qubit states.

For more than one qubit, a so-called computational basis of states is defined by
a product space 
\be
 \mid n \rangle_{n_q} \equiv   \mid q_1\rangle  \cdots  \mid q_{n_q}\rangle \equiv \mid \bf{Q} \rangle  ,
\ee  
where $n_q$ denotes the  total number of qubits in the system.  
We use the convention that the most significant 
qubit~\footnote{ An important aspect of relating the individual 
qubit state to a binary representation is that one can maintain 
the order of the qubits,  since if a qubit hops over to another 
order the decimal number is altered.} is labeled as $q_1$ and 
the least significant qubit by $q_{n_q}.$  Note we use $q_{i}$ to 
indicate the quantum number of the $i$th  qubit. The values assumed 
by any qubit are limited to either $q_i = 0$ or $1.$  The state label 
$\bf{Q}$  denotes the qubit array ${ \bf{Q}} = \left(  q_1, q_2, \cdots, q_{n_q} \right) ,$
which is a binary number label for the state with equivalent decimal label $n.$ 
This decimal multi-qubit state label is related to the equivalent binary label by 
\be
 n  \equiv  q_1 \cdot  2^{n_q -1}+ q_{2} \cdot   2^{n_q -2} + \cdots+  q_{n_q} \cdot   2^{0} = \sum_{i=1}^{n_q} \, q_i  \cdot   2^{n_q -i} \, .
\label{staten}
\ee  
Note that the $i$th qubit contributes a value of $  q_i  \cdot   2^{n_q -i}$
to the decimal number $n.$  Later we will consider ``partner states" 
($ \mid { n_0}   \rangle,\   \mid { n_1} \rangle $) associated with a 
given ${ n}, $ where a particular qubit  $i_s$  has a value of $q_{i_s} = 0,$
\be
{ n_0} =  { n} - q_{i_s}   \cdot   2^{n_q -i_s},
\label{pair0}\ee or a value  of
$q_{i_s} = 1,$
\be
{ n_1} =  { n} -  (q_{i_s}-1)   \cdot   2^{n_q -i_s}.
\label{pair1}
\ee  
These partner states are involved in the action of a single operator 
acting on qubit  $i_s ,$ as described in the next section.

A general state with $n_q$ qubits can be expanded in terms of the 
above computational basis states as follows
\be
 \mid \Psi\rangle_{n_q} = \sum_{ \bf Q} C_{ \bf Q} \mid { \bf Q} 
\rangle \equiv   \sum_{ n=0}^{2^{n_q}-1 }  C_{ n} \,  \mid  n \rangle\,,
\ee  
where the sum over ${ \bf Q}$ is really a product of $n_q$ summations 
of the form $\sum_{q_i=0,1}.$ The above Hilbert space expression maps over 
to an array, or column vector, of length $2^{n_q}$
\bea
  \qquad  \mid \Psi\rangle_{n_q}&\equiv& \left(
\begin{array}{l}
 C_0 \\
 C_1 \\
\ \, \vdots \\
\ \, \vdots \\
  C_{2^{n_q} -1}
\label{ampn}\end{array}
\right)   
\qquad
{\rm or\ with\ binary\ labels}
 \longrightarrow  
\left(
\begin{array}{lc}
 C_{0 \cdots 00} \\
 C_{0 \cdots 01} \\
\ \, \vdots \\
\ \, \vdots \\
 C_{1 \cdots 11}
\end{array}\right) \, .
 \eea    
The expansion coefficients $C_n$ (or $C_{{\bf Q}}$)  are complex numbers with
the physical meaning that $C_n=\langle n \mid \Psi\rangle_{n_q}$ is the
probability amplitude for finding the system in the computational basis state 
$\mid n\rangle, $ which corresponds to having the qubits pointing in the
directions specified by the binary array ${\bf Q}.$

Switching between decimal $n$ and equivalent binary ${\bf Q}$ labels 
is accomplished by the simple subroutines bin2dec and dec2bin in \QCMPI.
There we denote the binary number by an array $B(1) \cdots  B(n_q).$  The 
routines are: 

\begin{center}
\fbox{ \bf call bintodec(nq,B,D) } \quad
\fbox{ \bf call dectobin(nq,D,B) }
\end{center}

where {\bf nq} is the number of qubits; {\bf D} is a real decimal number and {\bf B} is the 
equivalent binary array.


\section{ONE-QUBIT OPERATORS}
\label{sec3}

An important part of quantum computation  is the act of rotating a qubit. 
The \NOT and single qubit Hadamard ${\bf \cal{ H} }$ operators are of 
particular interest:
\be
\NOT  \equiv \ \ \   \sigma_x  \longrightarrow  \left(
\begin{array}{lccr}
0&&& 1\\
1&&& 0
\end{array}\right) \, ,  \qquad
 {\bf \cal{ H} }  \equiv\ \ \   \frac{ \sigma_x + \sigma_z}{\sqrt{2}}
 \longrightarrow \frac{1}{ \sqrt{2}}  \left(
\begin{array}{lccr}
1&&& 1\\
1&&& -1
\end{array}\right)   \, .
\label{NOTH}  
\ee    
These have the following effect on the basis states 
$\NOT \mid 0\rangle  = \mid 1\rangle$, $\NOT~\mid~1~\rangle~=~\mid 0~\rangle$, and  
${\bf \cal{ H} }  \mid 0\rangle = \frac{ \mid 0\rangle +  \mid  1\rangle}{\sqrt{2}}\,,
 {\bf \cal{ H} }  \mid 1\rangle = \frac{ \mid 0\rangle -  \mid 1\rangle}{\sqrt{2}}.$
 
General one-qubit operators can be constructed from the Pauli operators; 
we denote the general one-qubit operator acting on qubit $s$ as $ {\Omega_s
}.$ Consider the action of such an operator on the multi-qubit state 
$\mid \Psi\rangle_{n_q} : $
\bea
 {\Omega_s }\!\!  \mid \Psi\rangle_{n_q}&=& \sum_{{\bf Q} }    C_{{\bf Q} } \  \      {\Omega_s }\!
 \!  \mid  {{\bf Q}}  \rangle \\ \nonumber
 &=&
\sum_{q_1=0,1}   \cdots  \sum_{q_s=0,1}   \cdots   \sum_{q_{n_q}=0,1}\ C_{{\bf Q} } \  \      
\mid  q_1 \rangle  \cdots \  (  {\Omega_s }\!\! \mid  q_s  \rangle )\    \cdots  \mid q_{n_q} \rangle. \\
\label{op1}
\eea  
  
Here $ {\Omega_s }$ is assumed to act only on the qubit $i_s$ of value $q_s.$  
The $  {\Omega_s }\!\! \mid  q_s  \rangle $ term can be expressed as
\be
{\Omega_s }\!\! \mid  q_s  \rangle =  \sum_{q'_s=0,1}\!\!   \mid q'_s \rangle \langle q'_s \mid\! {\Omega_s }\! \mid  q_s  \rangle,
\label{closure1} 
\ee  
using the closure property of the one qubit states. 
Thus Eq.~(\ref{op1}) becomes
\bea  
{\Omega_s} \!\!  \mid \Psi\rangle_{n_q}&=& \sum_{\bf  Q}   C_{\bf Q}\   {\Omega_s}  \!\! \mid  { \bf Q } \rangle
=  \\   \nonumber
\sum_{q_1=0,1}   \cdots  \sum_{q_s=0,1}   \cdots \!\!   \sum_{q_{n_q}=0,1}  \sum_{q'_s=0,1} \!\! C_{\bf Q} \!\!    
&&\langle q'_s \mid \!{\Omega_s }\! \mid  q_s  \rangle\  \mid  q_1 \rangle  
\cdots \   \mid  q'_s  \rangle  \cdots  \mid q_{n_q} \rangle.
\eea  
Now we can interchange the labels $ q_s \leftrightarrow  q'_s ,$  and use the 
label $ {\bf Q} $ to obtain the algebraic result for the action of a one-qubit 
operator on a multi-qubit state
\be
 {\Omega_s }  \mid \Psi\rangle_{n_q}= \sum_{\bf  Q}   {\tilde C}_{\bf Q}\   \mid  { \bf Q } \rangle =
 \sum_{n=0}^{ 2^{n_q}-1  }    {\tilde C}_{n}\   \mid n\rangle,
\ee    
where
\be
    {\tilde C}_{\bf Q}=  {\tilde C}_{n} =    \sum_{q'_s=0,1}\!\! \langle q_s \mid \!{\Omega_s }\! \mid  q'_s  \rangle\ 
   C_{\bf Q' ,} \label{oneopres}
\ee  
where  $ { \bf{Q}} = \left(  q_1, q_2, \cdots   q_{n_q} \right) ,$ and  
$ { \bf{Q'}} = \left(  q_1,\cdots  q'_{s} \cdots  q_{n_q} \right) .$  
That is ${ \bf Q}$  and ${ \bf Q'}$ are equal except for the qubit acted 
upon by the one-body operator  ${\Omega_s}.$  
    
A better way to state the above result is to consider Eq.~(\ref{oneopres})
for the case that $n$ has $q_s=0$ and thus $n\rightarrow n_0$ and to write 
out the sum over $q'_s$ to get
\be
{\tilde C}_{n_0} = \langle 0 \mid \!{\Omega_s }\! \mid  0  \rangle C_{n_0}+\langle 0  \mid \!{\Omega_s }\! \mid  1  \rangle C_{n_1},
\ee  
where we introduced the partner to $n_0$ namely $n_1.$ For the case that $n$
has $q_s=1$ and thus $n\rightarrow n_1$ Eq.~(\ref{oneopres}), with expansion 
of the sum over $q'_s$ yields 
\be
{\tilde C}_{n_1} = \langle 1 \mid \!{\Omega_s }\! \mid  0  \rangle C_{n_0}+\langle 1
\mid \!{\Omega_s }\! \mid  1  \rangle C_{n_1} \ , 
\ee 
 or,  written as a matrix equation,  we have for each $n_0, n_1$ partner pair
\be
  \left(
\begin{array}{l}
  {\tilde C}_{n_0} \\
  {\tilde C}_{n_1}
\end{array}
\right)   
=  \left(
\begin{array}{lccr}
\langle 0 \mid {\Omega_s} \mid  0  \rangle&&& \langle 0 \mid{\Omega_s} \mid  1  \rangle\\
\langle 1 \mid {\Omega_s} \mid  0  \rangle&&&\langle 1 \mid {\Omega_s} \mid  1  \rangle
\end{array}  \right) 
 \left(
\begin{array}{l}
 C_{n_0} \\
 C_{n_1}
\end{array}
\right) \ . 
 \label{resmat1} 
\ee 
This is not an unexpected result. Later we will denote the matrix element  
$\langle 0 \mid {\Omega_s} \mid  0  \rangle$ as $ {\Omega_s}_{0 0}$, etc.  

Equation~(\ref{resmat1})  above  shows how a $2\times2$ one-qubit operator 
${\Omega_s}$ acting on qubit $i_s$ changes the state amplitude for each value 
of $n_0.$  Here, $n_0$ denotes a decimal number for a computational basis
state with qubit $i_s$ having the $q_s$ value zero and $n_1$ denotes its
partner decimal number for a computational basis state  with qubit $i_s$ 
having the $q_s$ value one.  They are related by
\be
  n_1 = n_0 + 2^{n_q-i_s}.
\ee   
At times, we shall call $ 2^{n_q-i_s}$ the ``stride" of the $i_s$ qubit; 
it is the step in $n$ needed to get to a partner. There are $2^{n_q}/2$ 
values of $n_0$ and hence $2^{n_q}/2$ pairs $n_0,n_1.$  Equation
~(\ref{resmat1}) is applied to each of these pairs. In \QCMPI\ that process 
is included in the subroutine {\bf OneOpA}.  
  
Note that we have replaced the full $2^{n_q}\times2^{n_q}$ one qubit 
operator by a series of  $2^{n_q}/2$ sparse matrices. Thus we do not have 
to store the full  $2^{n_q}\times2^{n_q}$  but simply provide a $2\times2$
matrix for repeated use.  Each application of the $2\times2$ matrix 
involves distinct amplitude partners and therefore the set of $2\times2$ 
operations can occur simultaneously and hence in parallel. 
  
In the next section, the procedure for distributing the state vector over 
several processors is described along with the changes induced by the 
action of a one-body operator. Later this procedure is generalized to 
multi-qubit operators, using the same concepts.

\subsection{Distribution of the State}
   
The state of the multi-qubit system is described at any given time by 
the array of coefficients $C_n(t)$  for $n=0,\cdots 2^{n_q}-1,$  see  
Eq.~(\ref{ampn}).  The action of a one-qubit gate, assumed to act
instantaneously, is specified by the rules discussed in the previous 
section. Now we wish to distribute these state-vector coefficients stored 
in ``standard order" with increasing $n$, over a number of processors 
$N_P.$  For convenience, we assume that the number of processors invoked 
is a power of two, i.e. $N_P=2^p$ and thus we can distribute the $C_n$ 
coefficients uniformly over those processors with 
$N_x=2^{n_q}/N_P = 2^{n_q - p}$ amplitudes on each processor. In the code we 
denote $N_x$ as {\bf NPART}.
So, for example, we place  
\bea
&& C_0  \cdots  C_{N_x-1}\ \ \  \ \ \  \ \ \  \ \ \ \ \ \  \ \ \     {\rm  on\ processor \ 
   myid=0;} \\   \nonumber 
   && C_{N_x}  \cdots  C_{ 2 N_x-1} \ \ \ \ \ \  \ \ \  \ \ \  \ \ \  {\rm  on\  processor\  
   myid=1;}  \\  \nonumber
   &&  \ \ \ \ \ \ \ \  \vdots   \\  \nonumber
      &&  C_{ (N_P-1) N_x}  \cdots  C_{ N_P N_x-1} 
 \ \ \  \ \    {\rm  on\ processor\ 
   myid=N_P-1 \  . } 
\eea 
Where {\bf myid} is the processor number, from 0 to $N_p-1$. Note that 
$ N_P \cdot N_x=2^{n_q}.$   This distribution of the state over the
$N_P$ processors places a demand of $2^{n_q -p}$ on the memory of each
processor. For 64 processors $p=6$ and the memory required is down by a 
factor of 64; and for 4096 processors  $p=12$ and the memory required is 
down by a factor of 4096, etc. As the number of processors available 
increases, so will the memory demands on each processor.
   
However, life is not that simple. A one-qubit operator for a given partner 
pair $n_0,n_1$ often involves a $C_{n_0}$  that is seated on one processor and 
a $C_{n_1}$  that is seated on another  processor. We need to deal with that 
situation, while respecting the scheme for standard order distribution of the 
amplitude coefficients.  The first question that arises is when are the pairs 
$C_{n_0},C_{n_1}$ seated on the same processor?   We call that being ``seated 
in the same section," in analogy to theater seating.  That is, we dub being 
located on a particular processor as having the same section,  with the
location of a particular amplitude within that section being called its ``seat."  
With that language,  it is simple to state the condition for an amplitude 
pair  $C_{n_0},C_{n_1}$ being on the same processor; namely, that the
difference (we call this the ``stride") $n_0-n_1= 2^{n_q - i_s}$ be less than 
the distance  $2^{n_q }/N_P=2^{n_q - p}$  or simply  $i_s>p.$ If this
condition is not satisfied,  the stride is large enough to jump out of the 
section and thus require inter-processor communications. This result holds 
true if the number of processors is of the form $N_P=2^{ p}= 1,2,4,8,16
\cdots.$  One can prove this rule by induction.
   
This condition $i_s>p$, indicates that the larger $i_s,$ that is for qubits 
that are the least significant contributors to the state label $n,$  the 
associated pairs of amplitudes reside on the same processor. In contrast, 
the smaller $i_s$ are the qubits for which the pair amplitudes are the
furthest away in processor number. The stride ranges from a value of 1 for 
$i_s=n_q$ (least significant qubit) to $2^{n_q}/2$ for $i_s=1$ ( most
significant qubit). Carrying out the $2\times2$  matrix multiplication 
Eq.~(\ref{resmat1}) is simple for those pairs on the same processor,  but 
suitable  transfer to the requisite processors  must be implemented before 
one can perform the requisite  $2\times2$ matrix multiplication.  To carry 
out that process requires a way to identify the processor (e.g.  the section 
assignment) and the location within that processor (the seat ) and to 
interchange  the amplitudes.  The latter task is carried out using the 
MPI protocol, as discussed later.

\subsection{Pair Section, Seat and MPI}
  
Distribution of the $2^{n_q}$ amplitudes $C_0  \cdots  C_{2^{n_q} -1 }$ over 
the $N_P$ processors, places $N_x=2^{n_q}/N_P = 2^{n_q - p}$ amplitudes on 
each processor. As the state label $n$ ranges from $0$ to $2^{n_q} -1$  one 
jumps between different processors.  The relationship between the $n$ label 
and the processor on which  the associated amplitudes sits is simply:
${\rm  section}= {\rm Int}(n/N_x ),$  where Int() means the integer part 
and the seat (i.e. location within that processor) is 
${ \rm seat}={\rm Mod}(n,N _x)$   which denotes modular arithmetic of base 
$N_x.$  In the code $N_x$ is called {\bf NPART} and section is identical with
{\bf myid},  the processor number.

With the ability to identify the processor/location or section/seat 
assignment associated with each index $n,$ the remaining task is to 
transfer the requisite amplitudes to the ``correct" location. That task 
is carried out by the Message Passing Interface( MPI ) commands 
{\bf MPI\_SEND} and  {\bf MPI\_RECV}.  We need to coordinate the various 
processors  and exchange data during a calculation. The main reason MPI 
was developed over the last several decades is to enable efficient 
communication between processors during a computation.

Why use MPI? The MPI protocol affords many advantages for 
developing parallel processing codes. The main advantages are that: (1) MPI provides a standard set of  routines that are easy to use and (2) 
 MPI is flexible and works on many platforms.~\footnote{ We have 
run our codes on the Pittsburgh  and Barcelona supercomputers,  and also on arrays of imacs.}
  Thus MPI proved perfect for our need to develop a
user-friendly, flexible realization of the action of multi-qubit operators on 
state vectors  in a parallel computing environment.

\subsection{Action of  One-Qubit Operator}

  The following figures (Figs.~\ref{fig1}-~\ref{MPIrole2}) illustrate  the role played by MPI in transferring distributed amplitudes to appropriate processor
locations  when the one-qubit operator acts.   We use the case of $n_q=3$  or $2^3=8$ components as a simple example. 

The first case takes the partner labels $n=1, 3,$ which corresponds to the 
binary numbers $(001) $ and $(011).$  Here we use the binary labels for 
the components and consider the special case of a one-qubit operator acting 
on {\bf qubit 2} and assuming two processors $p=1$ (see Fig.~\ref{fig1} ). 
For that case, the two amplitudes affected by the one-qubit operator reside 
on the same processor, i.e., they have just different seats in the same section.
Thus there is no need for MPI data transfer.

\begin{figure}[t]
\begin{center}
\includegraphics[width=12cm]{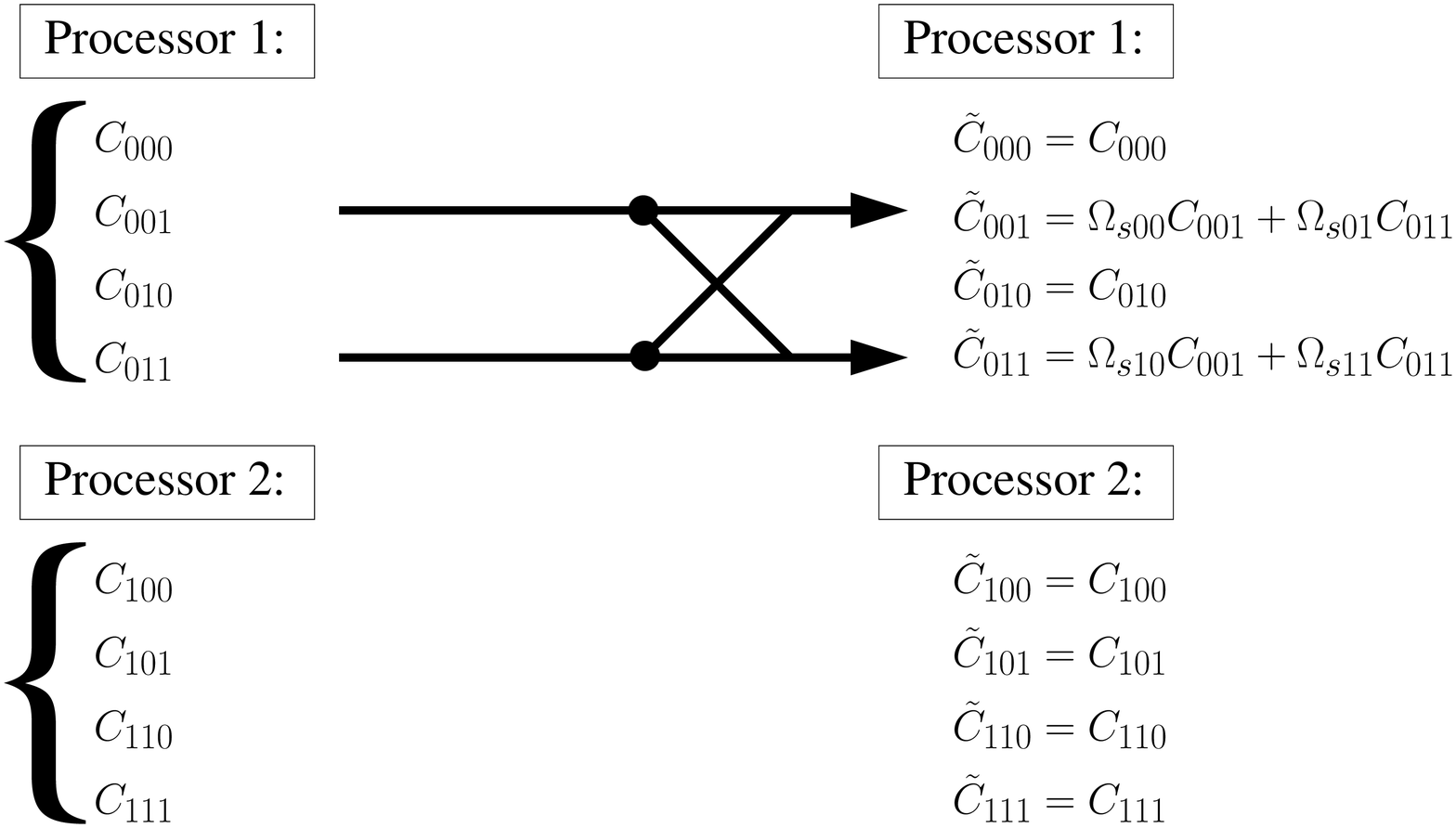}
\caption{A three qubit state vector is acted on by a one-qubit operator 
on qubit 2 ($i_s=2$) .  The case illustrated is for  the partners $n=1,3,$ 
which correspond to the binary numbers $(001) $ and $(011).$  It is assumed 
that there are just two processors $N_P=2^p,$ with $p=1.$
Thus $i_s>p$ for this case and the two coupled amplitudes reside on the same 
processor and no  MPI data transfer is invoked.  }
\label{fig1}
\end{center}
\end{figure}

  Now consider the partner labels $n=0,  4,$ which correspond to the binary 
numbers $(000) $ and $(100).$  Again we use the binary labels for the 
components, but now consider the special case of a one-qubit operator 
acting on {\bf  qubit 1} and  again assuming two processors. For this 
case, the two amplitudes affected by the one-qubit operator do not  
reside on the same processor, i.e., they are in different sections.
Thus there is now an essential need for MPI data transfer,  which 
involves sending and receiving as illustrated in Fig.~\ref{MPIrole2}.  
This entails two sends and two receives.

\begin{figure}[t]
\begin{center}
\includegraphics[width=12cm]{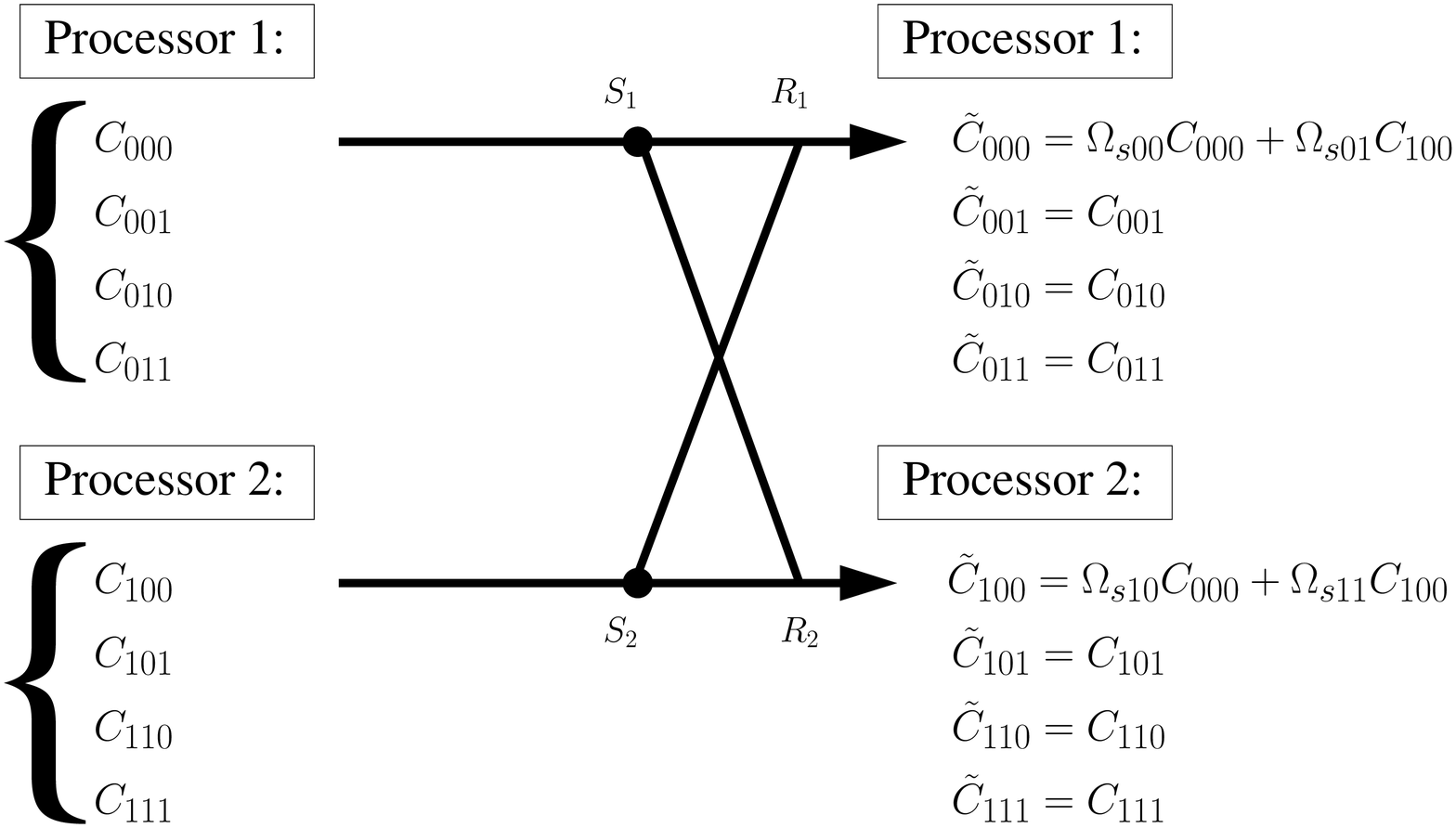}
\caption{A three qubit state vector is acted on by a one-qubit operator on 
qubit 1 $(i_s=1)$ . The case illustrated is for  the partners $n =0, 4,$ which 
corresponds to the binary numbers $(000) $ and $(100).$  It is assumed that 
there are just two processors $N_P=2^p,$ with $p=1.$ Thus the condition
$i_s>p$ is not satisfied and indeed the two coupled amplitudes reside on 
different processors and  MPI data transfer is invoked.  We need to send
$(S_1)$ component $C_{000}$ to processor one,  and it is received  at 
$(R_1),$   and also send $(S_2)$ component $C_{100}$ to processor zero, and 
it is received  at $(R_2).$  Later we will specify the send and receive commands in the MPI language. }
 \label{MPIrole2}
\end{center}
 \end{figure}

 Of course,  one needs to continue this process for the other three amplitude pairs
$n=1, 5,$ $n=2,6,$   and $n=3,7.$    In general,  we have $2^{n_q}/2$ partner pairs.\
Those pairs require six more sends and six more receives.
An important issue is then to see if the time gained by invoking more processors wins 
out over the time needed for all of these MPI transfers.  Another important concept is
one of ``balance," which involves the extent to which the various processors
perform equally in time and storage (ideally we assume they are all exactly 
equivalent and in balance). 

It is important to understand the above illustrations,  because for more
qubits and  more processors and for two- and three- qubit operators, the 
steps are  simply  generalizations of these basic cases. Careful manipulation 
of the amplitudes,  allows for spreading the amplitudes over many processors
and using MPI to do the requisite data transfers for all kinds of operators and gates.

For the one qubit case,  the steps here are called by the command  

\begin{center}
   \fbox{ \bf call OneOpA(nq,is,Op,psi,NPART, COMM) }
\end{center}   

where nq  denotes the number of qubits;  $"is''$ labels the qubit acted on by operator ``Op,"  psi is an
input wave function array of length {\bf NPART}=$N_x,$ which is returned as the
modified state vector. The last entry {\bf COMM} is included to allow for later 
extension to separate communication channels that we refer to as parallel universes or multiverses. 

Let us emphasize that any  operator, acting on one qubit is a 
special case of the one described here. Thus all rotations, 
all so-called local operations, including those generated by 
the one-body part of Hamiltonian evolution, are covered by 
the code {\bf OneOpA}.

 \section{MULTI-QUBIT OPERATORS}
\label{sec4}

Let us return to the main issue of how to distribute the amplitudes over
several processors and to cope with the action of operators on a quantum 
state.  The case of a two-qubit operator is a generalization of the steps 
discussed for a one-qubit operator.  Nevertheless, it is worthwhile to 
present those details, as a guide  to those who plan to use and perhaps 
extend \QCMPI.

We now consider a general two-qubit operator that we assume acts on 
qubits $i_{s_1}$ and $i_{s_2},$  each of which ranges over the full 
${1,  \cdots, n_q}$ possible qubits. General two-qubit operators can 
be constructed from tensor products of two Pauli  operators;  we denote the 
general two-qubit operator as $ {\bf \cal{ V} }.$   Consider the action of 
such an operator on the multi-qubit state $\mid \Psi \rangle_{n_q} : $
 \bea
 {\bf  \cal{ V} }\!\!  \mid \Psi\rangle_{n_q}&=& \sum_{{\bf Q} }    C_{{\bf Q} } \  \      {\bf \cal{ V} }\!
 \!  \mid  {{\bf Q}}  \rangle   \\ \nonumber
 &=&
\sum_{q_1=0}^{1}   \cdots  \sum_{q_{s1},q_{s2}=0}^{1}    \cdots  \sum_{q_{n_q}=0}^{1}\ C_{{\bf Q} } \  \        
\mid  q_1 \rangle  \cdots  (  {\bf {\cal V} }\!\! \mid  q_{s1}  q_{s2}   \rangle )\    \cdots  \mid q_{n_q} \rangle.
\label{op2}
\eea  
  
Here $ {\bf \cal{ V} }$ is assumed to act only on the two  $i_{s1},i_{s2}$
qubits. The $(  {\bf {\cal V} }\!\! \mid  q_{s1}\    q_{s2}   \rangle  )$ term 
can be expressed as
\be
{\bf {\cal V} }\!\! \mid  q_{s1}\  q_{s2}   \rangle  
=  \sum_{q'_{s1},q'_{s2}=0}^{1}  \mid q'_{s1}\    q'_{s2} \rangle \langle q'_{s1}\  q'_{s2}  \mid\! {\bf {\cal V} }\!\mid  q_{s1}\   q_{s2}   \rangle 
 \label{closure2} 
\ee  
using the closure property of the two-qubit product states. Thus Eq.~(\ref{op2}) becomes
\bea  
 {\bf { \cal V} } \!\!  \mid \Psi\rangle_{n_q}&=& \sum_{\bf  Q}   C_{\bf Q}\   {\bf {\cal V}}  \!\! \mid  { \bf Q } \rangle
= 
\sum_{q_1=0}^{1}   \cdots  \sum_{q_{s1}=0}^{1}   \cdots     \sum_{q_{s2}=0}^{1}   \cdots    \sum_{q_{n_q}
=0}^{1}  \sum_{q'_{s1},q'_{s2}=0}^{1}    \\   \nonumber C_{\bf Q}    
&&\langle q'_{s1} q'_{s2}  \mid \!{\bf {\cal V} }\! \mid  q_{s1} q_{s1}  \rangle\  \mid  q_1 \rangle  \cdots \   \mid  q'_{s1}     q'_{s2}  \rangle \cdots \   \mid q_{n_q} \rangle.
\eea  
Now we can interchange the labels $ q_{s1} \leftrightarrow  q'_{s1} , q_{s2}
\leftrightarrow  q'_{s2} $ and use the label $ {\bf Q} $ to obtain the
algebraic result for the action of a two-qubit operator on a multi-qubit state

\be
 {\bf  \cal{ V} }  \mid \Psi\rangle_{n_q}= \sum_{\bf  Q}   {\tilde C}_{\bf Q}\   \mid  { \bf Q } \rangle =
 \sum_{n=0}^{ 2^{n_q}-1  }    {\tilde C}_{n}\   \mid n\rangle,
   \ee    where
   \be
    {\tilde C}_{\bf Q}=  {\tilde C}_{n} =    \sum_{q'_{s1},q'_{s2}=0}^{1}\!\!
     \langle q_{s1}  q_{s2} \mid \!{\Omega_s }\! \mid  q'_{s1}  q'_{s2}  \rangle\ 
   C_{\bf Q' ,}
   \label{twoopA}
\ee  
where  $ { \bf{Q}} = \left(  q_1, q_2, \cdots   q_{n_q} \right) ,$ and  
$ { \bf{Q'}} = \left(  q_1,\cdots  q'_{s1} \cdots   q'_{s2} \cdots q_{n_q}
\right).$ That is ${ \bf Q}$  and ${ \bf Q'}$ are equal except for the qubits 
acted upon by the two-body operator  ${\bf {\cal V}}.$  
    
A better way to state the above result is to consider Eq.~(\ref{twoopA}) for
the following four choices 
\bea
n_{ 00} & \rightarrow &  ( q_1 \cdots   q_{s1}=0  \cdots  q_{s2}=0 ,\cdots q_{n_q} ) 
\nonumber \\
n_{01}   &  \rightarrow  & ( q_1 \cdots   q_{s1}=0  \cdots  q_{s2}=1 ,\cdots   q_{n_q} ) 
\nonumber \\
n_{10}   & \rightarrow &  ( q_1 \cdots   q_{s1}=1  \cdots  q_{s2}=0 ,\cdots q_{n_q} ) 
\nonumber \\
n_{11}   & \rightarrow &  ( q_1 \cdots   q_{s1}=1  \cdots  q_{s2}=1 ,\cdots   q_{n_q} ) , 
\label{twoopB}
\eea
where the computational basis state label $n_{q_{s1},q_{s2}}$ denotes the four 
decimal numbers corresponding to ${\bf Q} = ( q_1, \cdots q_{s1} \cdots  q_{s2 }  \cdots  q_{n_q}).$  
     
Evaluating Eq.~(\ref{twoopA}) for the four choices Eq.~(\ref{twoopB}) and  
completing the sums over $q'_{s1}, q'_{s2} ,$  the effect of a general
two-qubit operator on a multi-qubit state amplitudes is given by a $4 \times 4$ matix
    
 \be
  \left(
\begin{array}{l}
  {\tilde C}_{n_{00}} \\
  {\tilde C}_{n_{01}} \\
  {\tilde C}_{n_{10} }\\
  {\tilde C}_{n_{11}}
\end{array}
\right)   
=  \left(
\begin{array}{lccccr}
 {\cal V}_{00;00} &&  {\cal V}_{00;01}  
 \ \ \  {\cal V}_{00;10}      \ \ \       {\cal V}_{00;11}\\
{\cal V}_{01;00} &&  {\cal V}_{01;01}  
 \ \ \  {\cal V}_{01;10}      \ \ \       {\cal V}_{01;11}\\
 {\cal V}_{10;00} &&  {\cal V}_{10;01}  
 \ \ \  {\cal V}_{10;10}      \ \ \       {\cal V}_{10;11}\\
{\cal V}_{11;00} &&  {\cal V}_{11;01}  
 \ \ \  {\cal V}_{11;10}      \ \ \       {\cal V}_{11;11}
\end{array}  \right) 
 \left(
\begin{array}{l}
 C_{n_{00}} \\
  C_{n_{01}} \\
  C_{n_{10} }\\
 C_{n_{11}}
\end{array} 
\right) \ ,
\label{resmat2} 
\ee   
where  $ {\cal V}_{ij;kl} \equiv  \langle i,j   \mid  {\cal V} \mid  k , l
\rangle.$ Equation~(\ref{resmat2})  above    shows how a $4\times4$ two-qubit operator 
${\cal V}$ acting on qubits  $i_{s1},i_{s2} $ changes the state amplitude for
each value of $n_{00}.$  Here, $n_{00}$ denotes a decimal number for a
computational basis state with qubits  $i_{s1},i_{s2} $ both having the
values zero and its three partner decimal numbers for a computational basis state
with qubits $i_{s1},i_{s2} $ having the values  $(0,1), (1,0)$ and $(1,1),$ 
respectively.  The four partners $n_{00},n_{01},n_{10},n_{11},$ or ``amplitude
quartet," coupled by the two-qubit operator are related by:
\be
n_{01}=n_{00} +  2^{n_q - i_{s2}} \qquad n_{10}=n_{00} +  2^{n_q - i_{s1}}
\qquad n_{11}=n_{00} +  2^{n_q - i_{s1}}+2^{n_q - i_{s2}},
\ee  
where $ i_{s2}, i_{s2}$ label the qubits that are acted on by the two-qubit operator.

There are $2^{n_q}/4$ values of $n_{00}$ and hence $2^{n_q}/4$ amplitude
quartets $n_{00}, n_{01}, n_{10},n_{11}.$ Equation~(\ref{resmat2}) is applied 
to each of these quartets for a given pair of struck qubits. In \QCMPI\   that
process is included in the subroutine TwoOPA.
  
In this treatment, we are essentially replacing a large sparse matrix, 
by a  $2^{n_q}/4$ set of  $4 \times 4 $ matrix actions, thereby saving 
the storage of that large matrix.
  
In the next section, the procedure for distributing the state vector 
over several processors is illustrated along with the changes 
induced by the action of a two-body operator. 

  \subsection{Action of  Two-Qubit Operators}

To visualize the distribution of the amplitudes over several processors and 
the role played by MPI in transferring the amplitudes to appropriate location, 
when the {\it two}-qubit operator acts,  the following diagrams lay out 
the scheme. We again use the case of $n_q=3$  or $2^3=8$ components as a simple illustration.  

The first case takes the amplitude quartet labels $n=0,1,2,3$ which
corresponds to the binary numbers $(000), (001), (010),$ and $(011).$  Here 
we use the binary labels for the components and consider the special case 
of a two-qubit operator acting on {\bf  qubits 2 and 3.}  We consider 
just two processors. In this case the four amplitudes affected by the 
two-qubit operator reside on the same processor, i.e., they have just 
different seats in the same section. Thus there is no need for MPI data transfer.
 \begin{figure}[t]
\begin{center}
\includegraphics[width=10cm]{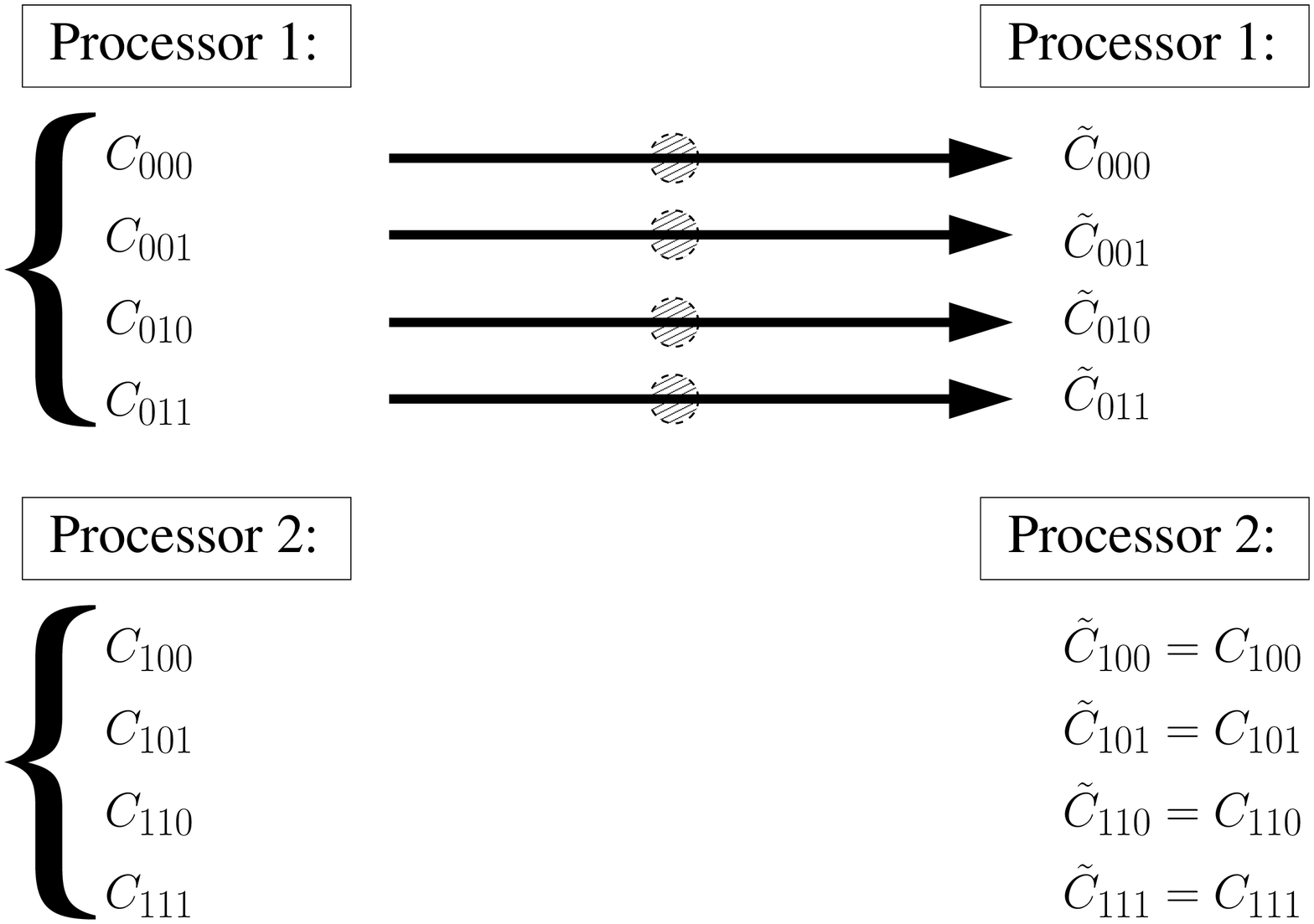}
\caption{A three qubit state vector is acted on by a two-qubit operator on
  qubits 2 and 3 ($i_{s1}=2,i_{s2}=3$) .  The case illustrated is for  the 
amplitude quartet $n =0,1,2,3, $ which corresponds to the binary numbers 
$(000) ,(001), (010),$ and $(011).$   It is assumed that there are just 
two processors $N_P=2^p,$ with $p=1.$ Thus $ i_{s2} > i_{s1}>p$ for this 
case and  the two coupled amplitudes reside on the same processor and 
no MPI data transfer is invoked. The dashed circles indicate that all four 
amplitudes contribute to forming the values of 
${\tilde C}_{000},   {\tilde C}_{001} , {\tilde C}_{010},  {\tilde C}_{011} $ 
are given by Eq.~(\ref{twoopA}).  }
 \label{MPIrole2A} 
\end{center}
\end{figure}

\begin{figure}[t]
\begin{center}
\includegraphics[width=10cm]{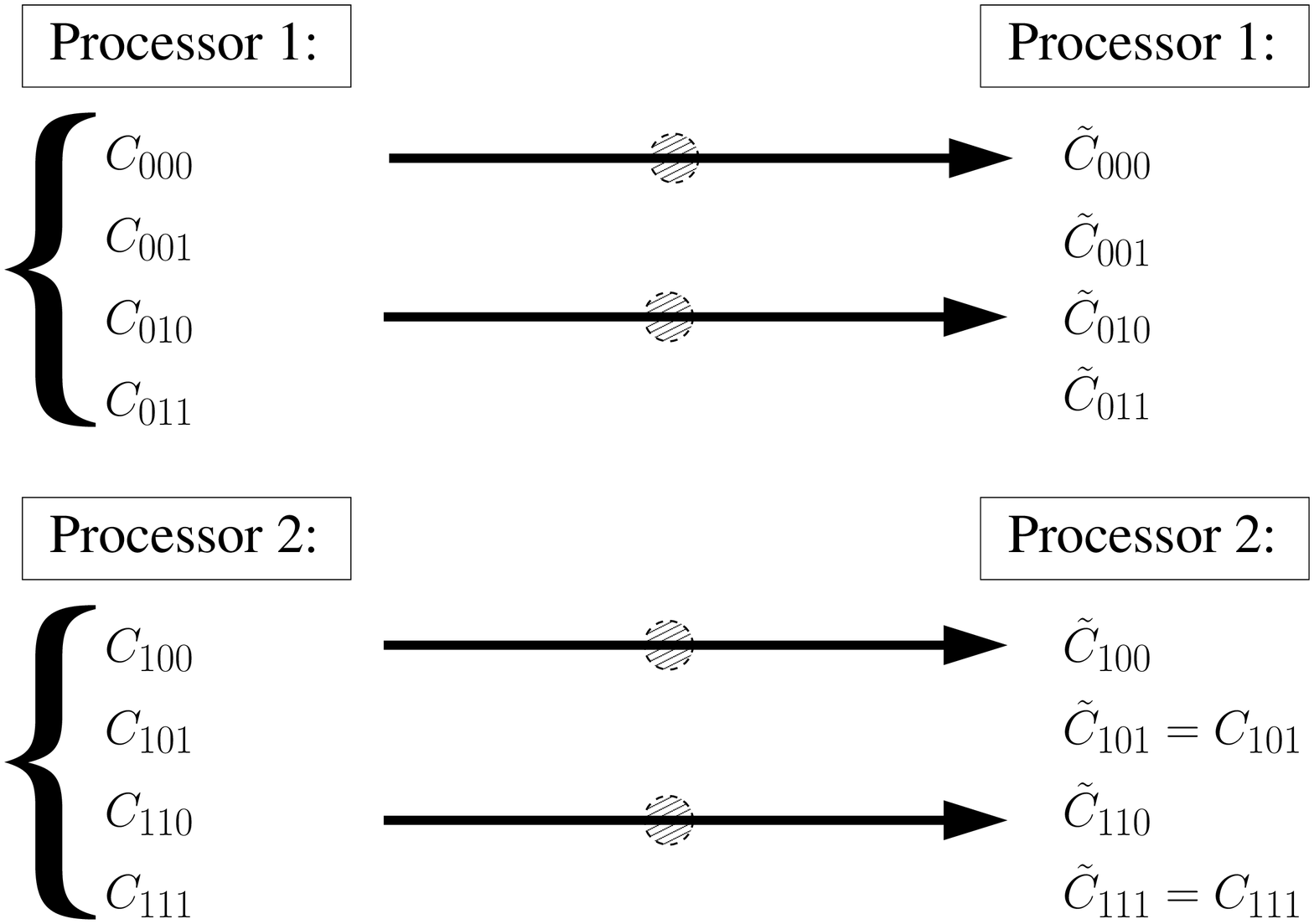}
\caption{A three qubit state vector is acted on by a two-qubit operator on 
qubits 1 and 2  ($i_{s1}=1,i_{s2}=2$) .  The case illustrated is for  the 
amplitude quartet $n =0,2,4,6, $ which corresponds to the binary numbers 
$(000) ,(010), (110),$ and $(110).$   It is assumed that there are just 
two processors $N_P=2^p,$ with $p=1.$ Thus $ i_{s2} > p,$ but we do {\bf not} 
have $i_{s1} >p,$ thus for this case amplitudes reside on different processors
and MPI data transfer is invoked. The dashed circles indicate that all four 
amplitudes are to be sent/received from other locations.  }
 \label{MPIrole2B}
\end{center} 
\end{figure}

  Now consider the amplitude quartet labels $n=0,  2 ,4,  6,$ which 
corresponds to the binary numbers  $(000), (010), (100),$ and $(110).$  
Again we use the binary labels for the components, but now consider 
the special case of a two-qubit operator acting on {\bf qubits 1 and 2.} 
  We consider just two processors.  For this case, the two amplitudes 
affected by the one-qubit operator do not  reside on the same processor, 
i.e., they are in different sections. Thus there is now an essential 
need for MPI data transfer, which involves sending and receiving
as illustrated in Fig.~\ref{MPIrole2B}.
  
For the two qubit case,  the steps here are called by the command  

\begin{center}
\fbox{ \bf call TwoOpA(nq,is1,is2,Op,psi,NPART,COMM) }
\end{center}  

where {\bf nq} denotes the number of qubits; {\bf is1,is2} label the qubits acted on by   
 operator {\bf Op},  {\bf psi} is an input wave function array of length {\bf NPART}=$N_x,$  which
 is returned as the modified state vector.

\subsection{CNOT and CPHASE}

The two-qubit operators {\bf CNOT} and {\bf CPHASE} are oft-used special cases 
of the above two-qubit operator discussion.  They are simpler than the general case because 
they are given  by the sparse matrices
\be
{\cal V}  \rightarrow   {\rm CNOT} = 
\left(
\begin{array}{lccr}
1 & \ \ 0 & \ \  0 & \ \   0\\
0 & \ \ 1 & \ \  0 & \ \  0\\
0 & \ \ 0 & \ \ 0 & \ \  1\\
0 & \ \ 0 & \ \ 1 & \ \  0\\
\end{array}\right)\, , \quad
{\cal V}  \rightarrow   {\rm CPHASE} = 
\left(
\begin{array}{lccr}
1 & \ \ 0 & \ \ 0 & \ \  0\\
0 & \ \ 1 & \ \ 0 & \ \  0\\
0 & \ \ 0 & \ \ 1 & \ \  0\\
0 & \ \ 0 & \ \ 0 & \ \  -1\\
\end{array}\right)  \, .
\ee  
{\bf CNOT} stores the rule $00 \rightarrow 00, 01 \rightarrow 01,10 \rightarrow 11,11 \rightarrow 10 ,$
where qubit 1 is the control,  and qubit 2 gets acted on by $\sigma_1$ 
only when qubit 1 has a value of one. In \QCMPI, a  subroutine {\bf CNOT} codes 
this special two-qubit operator:

\begin{center}
\fbox{ \bf call CNOT(nq,is1,is2,psi,NPART,COMM) }.
\end{center}

{\bf CPHASE} stores the rule 
$00 \rightarrow 00, 01 \rightarrow 01,10 \rightarrow 10,11 \rightarrow -11 ,$
where qubit 1 is the control,  and qubit 2 gets acted on by $\sigma_3$  only
when qubit 1 has a value of one. In \QCMPI, a  subroutine  {\bf CPHASEA} codes this 
special two-qubit operator:
 
\begin{center}
\fbox{ \bf call CPHASEA(nq,is1,is2,psi,NPART,COMM) }.
\end{center}

Another two-qubit operator which plays a key role in  the quantum Fourier transformation,
is the {\bf CPHASEK},  given by a sparse matrix that depends on a positive integer $k$
\be
{\cal V}  \rightarrow   {\rm CPHASEK} = 
\left(
\begin{array}{lccr}
1 & \  \  \ 0 & \  \  \ 0 &0 \ \  \\
0 & \  \  \ 1 & \  \  \ 0 &0 \ \  \\
0 & \  \  \ 0 & \  \  \ 1 &0 \ \  \\
0 & \  \  \ 0 & \  \  \ 0 & \  \ e^{ \frac{2  \pi i}{2^k}} \\
\end{array}\right)  \, .
\label{cphasek}
\ee  
In \QCMPI, a  subroutine  {\bf CPHASEK} codes this special two-qubit operator:

\begin{center}
 \fbox{ \bf call CPHASEK (nq,is1,is2,k, psi,NPART,COMM) }.
\end{center}

 Note that there are no MPI commands required in this subroutine.

\subsection{The Full Hadamard--Special Handling} 
\label{sec5C}

An important example of a multi-qubit operation is when  Hadamards act on all
of the qubits in a system--a step that is often  used to initialize a QC. 
One way to do that is simply to repeat the prior discussion
and use the subroutine  for qubits $i_s =1  \cdots n_q.$  That procedure is implemented in the subroutine {\bf HALL}.

     Hadamards acting on all qubits involves the operator   
\be
 {\cal H}^{n_q} \equiv  \left [ {\cal H}_1  \otimes{\cal H}_2  \cdots   \otimes {\cal H}_{n_q}\right]   .
\label{HALL}
\ee Another way to implement this operator is based on the realization
 that when acting on the column vector  $(C_{0}  \cdots  C_{2^{n_q} -1} )  $ it forms an equal
weighted combination with particular signs $s_{n.n'},$  whereby the effect of $ {\cal H}^{n_q} $ is
\be
C_n \rightarrow \frac{1}{2^{\frac{n_q}{2} }}  
\sum_{n'=0}^{2^{n_q} -1}s_{n,n'} C_{n'} \equiv {\tilde C}_n \,  . 
\ee   
The task is to determine the signs  $s_{n,n'}.$  
These signs are relatively simple to pin down.
From the product structure of  $ {\cal H}^{n_q} $ it is simple to show that the 
signs are determined by the condition $s_{n,n'}=(-1)^\delta,$  where $\delta$ 
denotes the number of times the binary bits for $n,n'$ of unit value are 
equal, i.e.  how many  times $B(i)=B'(i)=1.$  This condition is carried 
out in the  function  SH: 

\fbox{ \bf FUNCTION SH(nq,n,np) }
   
where nq  is the number of qubits; $np=n' .$  This procedure is implemented in the subroutine {\bf HALL2}.
The user should decide which version  {\bf HALL} or  {\bf HALL2} works best in their context.

Ironically, although a small subroutine,  {\bf HALL} or  {\bf HALL2}  is used repeatedly in 
Grover's search and  dominate the  time 
expended in a large qubit quantum search.  We shall refer to the operator $ {\cal H}^{n_q} $
as  {\bf HALL} throughout this paper, recognizing that it can be implemented using either 
 {\bf  CALL HALL} or by the special sign handling  method  {\bf CALL HALL2}. 

\section{A SAMPLE OF RELEVANT QUANTUM ALGORITHMS}

\QCMPI permits the simulation of any quantum circuit on a 
parallel computing environment. In this section we describe 
two well-known QC algorithms already included in the current 
package and which exemplify the use of \QCMPI in practical 
applications.

\subsection{GROVER's searching algorithm}
\label{sec5}

We now show how to apply the operators (gates) , and the  treatment of a 
multi-qubit state,  to the first of several basic QC algorithms.  These 
are standard procedures in QC and we examine them with \QCMPI\ so 
that one can describe these algorithms dynamically using basic, 
realistic Hamiltonians and also subject these procedures to environmental effects.
The case of superdense coding,~\cite{superdense} which  is a way to 
enhance communication between Alice and Bob by means of shared 
entangled states., has also been developed in \QCMPI.

Our first application presented here is Grover's search
algorithm~\cite{Grover}. In this case, we start with a state of $n_q$ qubits 
all pointing up $\mid 0 0 0 \cdots   0 \rangle,$ and  act on it with {\bf HALL}, 
see Eq.~(\ref{HALL}). Then, we need  an all-knowing Oracle operator
$\oracle$ to mark an item that is to be ferreted out by the algorithm.
The Oracle step is very simple when we use amplitude coefficients $C(n);$ 
we simply find the processor (section)  and location on that processor 
(seat) associated with the marked item $n_x$ and reverse the sign of that 
amplitude $C(n_x) \rightarrow - C(n_x).$  All other amplitudes are unchanged.
\be
\oracle 
 \left(
\begin{array}{l}
 C_{0} \\
  C_{1} \\
\   \vdots \\
   C_{n_x} \\
\    \vdots \\
  C_{2^{n_q}-1}  
\end{array} 
\right)  \longrightarrow
 \left(
\begin{array}{l}
 C_{0} \\
  C_{1} \\
\   \vdots \\
   -C_{n_x} \\
\    \vdots \\
  C_{2^{n_q}-1}  
\end{array} 
\right) \, .
\label{oracle1}
\ee
This process can be extended to two or more marked items.  

Grover's procedure,  which entails using constructive interference to 
make the marked item's amplitude stand out from all others, involves 
acting repeatedly on the state {\bf HALL}$ \mid 0 0 0 \cdots   0 \rangle$ 
with the ``Grover operator"
\be
{\cal G} \equiv {\rm HALL} \ \cdot \  {\cal Inv}\  \cdot \  {\rm HALL} \ \cdot
\ \oracle
\ee
where ${\cal Inv}$  is an operator
\be
 {\cal Inv} = 2   \mid 0 0 0 \cdots   0 \rangle  \langle 0 0 0 \cdots   0  \mid - {\bf  I}
\ee where ${\bf I}$ is an $2^{n_q}\times2^{n_q}$ unit matrix. The 
operator $ {\cal Inv}$ is simply realized by changing  the sign of 
all amplitudes except for the $n=0$ one.
\be
{\cal Inv} 
 \left(
\begin{array}{l}
 C_{0} \\
  C_{1} \\
   C_{2} \\
\   \vdots \\
  C_{2^{n_q}-1}  
\end{array} 
\right)  \longrightarrow
 \left(
\begin{array}{l}
+ C_{0} \\
 - C_{1} \\
  - C_{2} \\
\   \vdots \\
  -C_{2^{n_q}-1}  
\end{array} 
\right) \, .
\label{inversion}
\ee  The combination {\bf HALL} $ \cdot \ {\cal Inv} \ \cdot $ {\bf HALL} is called an 
inversion about the mean and is an essential part of Grover's algorithm. 
The other essential part is to act on the initial state enough times $n_t$ 
to produce an amplitude ${\bf {\tilde C}(n_x)}$ that stands out with  
high probability.  We take the number $n_t= \frac{\pi}{4}  \sqrt{ 2^{n_q }  }   $\ .
\be
G^{n_t}  \cdot {\rm HALL} \mid 0 0 0 \cdots   0 \rangle \rightarrow
 \left(
 \begin{array}{l}
 {\tilde C}_{0} \\
    {\tilde C}_{1} \\
   \  \vdots \\
   {\bf  {\tilde C}_{n_x}} \\
\   \vdots \\
  {\tilde C}_{2^{n_q}-1}  
\end{array} \, .
\right)
\ee
In all of these simple states,  it is the ${\rm {\bf HALL}}$ operator and its 
repeated application in $G^{n_t}$ that involves the most time expenditure.

The  \QCMPI\ Grover search for a large number of qubits is included as Guniv.f90. Instructions for
running the code and a guide to steps invoked are incorporated directly as comments in the listing.
Generalization to include noise using a multiuniverse approach is discussed later.

\subsection{SHOR's factoring algorithm}
\label{sec6}

Shor's algorithm~\cite{Shor} is a QC method for factoring a large number.
The basic idea is to prepare a state that, when subjected to a Quantum 
Fourier Transform (QFT),  permits one to search for the period of a function 
that reveals the requisite factors with high probability.  It uses quantum 
enhancement to go way beyond classical factoring procedures and yields the 
factors with high probability for very large non-prime numbers, after 
relatively few tries. To simulate this algorithm,  where we are restricted 
to numbers much smaller than possible in a future QC, there are several 
steps  implemented in the \QCMPI \  code  subshor.f90. A pedagogic analysis 
of the reasons for each step of Shor's algorithm is presented in reference~\cite{Gerjuoy} .
\subsection*{ {\bf Step 1:} Choose the number, $M,$ and set the register sizes}
Choose the number, $M,$ to be factorable number:  
$15, 21, 33, 35, 39,55,77 \cdots $ and determine the  size of two requisite 
registers.

Preparatory tests on  the input number $M$ are made so that the code 
continues only if the input number is not a power of 2 or a prime and 
is thus a suitable candidate for factoring.  These are classical procedures 
performed by standard codes.  Then, based on having an acceptable input 
number, an initial state of two distinct registers are prepared, with 
register one having $n_1$  and register two having $n_2$ qubits. The 
first register should ~\cite{Shor,Gerjuoy} have enough qubits to store 
in base 2 all numbers in the range  $M^2$  to $2 M^2,$  i.e. 
$ M^2 \geq 2^{n_1}< 2 M^2.$ Therefore the choice for $n_1$ is set as
\be
n_1 = {\rm Ceiling}(\log_2(Q)\  )  \qquad   Q=2^{{\rm Ceiling}(\log_2(M^2)\  )}
\ee 
where ${\rm Ceiling}( x)$ gives the smallest integer greater than or 
equal to $x.$~\footnote{   For example,  if  $M=21,$  and  $M^2=441,$ then 
$n_1=9$ and $2^9=512,$ and register one includes the value $441.$ If $n_1$ 
were taken as $8,$ then register one would not include the value $441,$  
since $2^8=256.$ If $n_1$ were taken as $10,$ then register one would exceed 
the value $2 M^2=882,$  since $2^{10}=1024.$    }

The number of qubits in register two is set by
\be
n_2 = {\rm Ceiling}(\log_2(M)\  ) 
\ee 
so that there are enough qubits to store in base 2 all numbers  up to and 
including the value of the input number   $M.$ ~\footnote{For example,  
if  $M=21,$ then $n_2=5$ and $2^5=32>M,$ and register two includes the 
value $21.$ If $n_2$ were taken as $4,$  then register two would not include 
the value $21,$ since $2^4=16.$}

Here we invoke  the minimum number of qubits for both registers. A larger 
$n_1$  lengthens the computation, albeit providing higher probability of success.

\subsection*{ {\bf Step 2 :} Load the first register}

Load the first register with all the integers less than or equal 
to $2^{n_1}-1$.  

This is achieved by acting with a Hadamard on all qubits in register one, that is 
use {\bf HALL} on a basis state of $n_1$ qubits so that register one is set to the state
\be
\mid \Psi_1\rangle  = {\rm HALL} \mid 00  \cdots  0\rangle_{n_1} 
= \frac{1}{2^{\frac{n_1}{2} }}\   \sum_{n=0}^{2^{n_1}-1}\ \mid n\rangle_{n_1}.
\label{Shor1}
\ee 
Thus each of the $2^{n_1}$ amplitudes appear with equal weight, which is the 
quantum massive parallel processing feature.

Next we attend to setting register two.

\subsection*{ {\bf Step 3 :} Load the second register}
Select an integer $({\rm xguess})$ coprime to M and load the function
$f(j) \equiv  {\rm xguess}^j \  (mod M)$  into the second register for a fixed 
choice of ${\rm xguess}$ and all possible $j$ values:  $ 0 \leq \, j \,  \leq 2^{n_1}-1.$

Note that in \QCMPI, we simply compute $\mid f(n)\rangle$  and tack it on to 
the $\mid n\rangle_{n_1}$ state. As discussed by Shor~\cite{Shor}, one must 
actually do this crucial step by a quantum process for modular exponentiation.

In Shor's algorithm, ${\rm xguess}$ is a random choice for a number that is coprime 
to M,  where coprime means that $M$ and ${\rm xguess}$ have no common factor other 
than 1.  Euler's phi function of $M$ is used to determine the range of 
integers between 1 and $M$ that are coprime to $M;$  a number in this 
interval  $a$ is selected and then tested using ${\rm GCD }[a,M]\equiv 1$ to 
assure that it is coprime to $M.$ If it passes that test we set the 
value for $a \rightarrow {\rm xguess}.$ 

The reason for defining the above function (aka the Shor Oracle) is that 
this function  $f(j) $has a characteristic period  for each value of $M$ and ${\rm xguess}.$  
\be
\mid f(n+r)\rangle=\mid f(n)\rangle  , \qquad  f(n)  = {\rm xguess}^n \  (mod M).
\label{fperiod}
\ee
 Finding the period $r$ is a key goal.

 The full state composed of registers 1 and  2 $(n_q=n_1+n_2)$ is built in the 
following way:
\be
\mid \Psi\rangle_{n_q}  = \frac{1}{2^{\frac{n_1}{2} }}\    
\sum_{n=0}^{2^{n_1}-1} \  \mid n\rangle_{n_1} \mid f(n) \rangle_{n_2}\,.
\label{Shor2}
\ee

 \subsection*{ {\bf Step 4 :} Measure register 2}
 
We could measure the second register next or postpone that act to 
coincide with step 6 below, because step 5 involves register 1 only.  It 
is helpful to think of the action of measuring register 2 now to motivate 
the need for step 6.   For each possible value of $n_1\rightarrow k$, one 
asks if register 2 is in state $ _{n_2}\, \!\!\langle k\mid$  by projecting the 
full state
 \be
 _{n_2}\!\!\langle k\mid \Psi\rangle_{n_q}  = \frac{1}{2^{\frac{n_1}{2} }}\ 
\sum_{n=0}^{2^{n_1}-1} \  \mid n\rangle_{n_1} \langle k\mid f(n)
\rangle_{n_2} \rightarrow \frac{1}{D^{\frac{1}{2} }}\    
\sum_{j=0}^{D-1} \  \mid n_k + j r\rangle_{n_1} ,
 \label{projreg2}
 \ee  
where at the last stage the state is normalized after projection--the usual 
Born rule for a projective measurement. Note that for every choice of 
$k,   D$  terms of the first register appear in  superposition  where  
$D$ is  $ \approx 2^{n_1} /r . $ The integer  $ D$ is 
ascertained~\footnote{ The value of $D$ is constrained by the conditions 
$ 0 \leq n_k + (D-1) r  \leq  2^{n_1}-1 $   and $ 0 \leq n_k  \leq  r-1.$  
Hence, the integer $D$ is constrained by 
$D \leq  \frac{ 2^{n_1}}{r} +\frac{r-1-n_k}{r}.$ }  by the period $r$ for a 
fixed ${\rm xguess}$ by
 \be
 \mid f(n_k+ (D-1) r)\rangle= \mid f(n_k+ (D-2) r)\rangle=\cdots=  
\mid f(n_k+  1 r)\rangle=\mid f(n_k)\rangle=k.
 \ee  
The next step involves acting on register 1 to search for the 
period $r$  by enhancing the quantum interference using a  quantum Fourier transform.

\subsection*{ {\bf Step 5 :} Quantum Fourier Transform register 1}

Register one, which is now in the state
\be
\mid \Phi\rangle_{n_1} =  \frac{1}{\sqrt{D}} \    \sum_{j=0}^{D-1} \  \mid n_k + j r \rangle_{n_1},
\label{Phistate}
\ee is next acted upon by a quantum Fourier transformation operator 
${\bf\rm QFT}$ (see the Appendix for a discussion of the QFT operator) which changes the above state
\bea
{\rm QFT} \mid \Phi\rangle_{n_1} &=&  \frac{1}{\sqrt{ 2^{n_1} D}} \     
\sum_{j=0}^{D-1} \    \sum_{n=0}^{2^{n_1}-1} \  
e^{ 2 \pi i n   (n_k + j r)/2^{n_1}}   \mid n \rangle_{n_1} \nonumber \\
&=&   \frac{1}{\sqrt{ 2^{n_1} D}} \     \sum_{n=0}^{2^{n_1}-1} \  
e^{ 2 \pi i n   n_k /2^{n_1}}    \sum_{j=0}^{D-1} \  (e^{ 2 \pi i n  r/2^{n_1}} )^j   \mid n \rangle_{n_1}\,. 
\label{QFT1}
\eea  The QFT is a unitary operator which switches to a basis in which the superposition
is isolated into the above exponential amplitude.  The sum on $j$ can be 
performed~\footnote{  The following summation rule is used here 
$ \sum_{j=0}^{D-1}\   X^j  =   \frac{X^D -1}{X -1}.$  One can also display
  this as  2-D vector additions of equal length phasors, as in  Fresnel zone
  plate interference.} and thus the result is
\be
{\rm QFT} \mid \Phi\rangle_{n_1} =     \frac{1}{\sqrt{ 2^{n_1} D}} \   
\sum_{n=0}^{2^{n_1}-1}  \  e^{ 2 \pi i n  n_k /2^{n_1} } \    e^{  \pi i n (D-1) /2^{n_1} }\  
\frac{ \sin( D  \pi  n  \frac{r}{2^{n_1}}  ) }{\sin(\pi  n \frac{r}{2^{n_1}}
  )}    
\mid n \rangle_{n_1}  .
\label{QFT2}
\ee  
The probability for finding the final state  with register 1 in state $
\langle n\mid $ 
and  register 2 in state $ \langle k \mid $ is therefore
\be
p(n,k ) =  \frac{1}{  2^{n_1} D} \  
\left[\frac{ \sin( D  \pi  n \frac{r}{2^{n_1}}  ) }{\sin(\pi  n \frac{r}{2^{n_1}} )} \right]^2,
\label{probshor} \ee  where the dependence on $k$ has dropped out, 
(except for a possible  dependence of $D$ on $k$-see earlier footnote).   
Note that for the special case that $ n  r/2^{n_1}$ is an integer, the 
above result reduces to $p(n,k ) \rightarrow  \frac{D}{  2^{n_1} } \approx
\frac{1}{r} , $  
which is understood by returning to Eq.~(\ref{QFT1}) and using 
$\sum_{j=0}^{D-1} \  (1 )^j  =D.  $

How can one extract  the period $r$ from making 
such a measurement on registers 1 and 2?

\subsection*{ {\bf Step 6:} Measure register 1 and determine period and factors}

The measurement of registers 1 and 2 has a probability given by
Eq.~(\ref{probshor}). At select values of $n\rightarrow \bar{n},$ the 
probability $p(n,k )$ has local maxima.  Consider the associated
fraction $\frac{ \bar{n} }{2^{n_1}},$  which is extracted from a 
determination of those local maxima. At these maxima 
\be
p_{max} =  \frac{1}{  2^{n_1} D} \    
\left[\frac{ \sin( D  \pi  r  \frac{\bar{n}}{2^{n_1}}  ) }
{\sin(\pi r  \frac{\bar{n}}{2^{n_1}} )} \right]^2.
\label{probshor2} 
\ee 
In the arguments of the sin functions in Eq.~(\ref{probshor2}) , $D$ is an 
integer,  so the maximum probability occurs in the vicinity of  an integer value for 
$r  \frac{\bar{n}}{2^{n_1}} .$  We therefore seek an approximate value of the 
ratio $ \frac{\bar{n}}{2^{n_1}} \approx integer/r,$ for an even $r.$  That 
ratio is found by expressing $  \frac{\bar{n}}{2^{n_1}} $ as a continued 
fraction and determining its first convergent of the form $integer/r,$ 
for an even $r.$\footnote{Gerjuoy~\cite{Gerjuoy} showed that the maximum 
probability is not less than $ \frac{4}{ \pi^2} \approx 0.4,$  but more 
likely to be $\geq  \frac{8}{ r \pi^2} \approx 0.81$}

 That determines the value of the period $r,$  which we require to be even 
so that we can use the final step~\footnote{ Note that the periodic function 
${\rm xguess}^r {\rm Mod}[M] = {\rm xguess}^0 {\rm Mod}[M] =1.$   For even period $r$ this yields 
$ ({\rm xguess}^{\frac{r}{2}})^2 -1 \equiv 0\  {\rm Mod}[M]= ({\rm xguess}^{\frac{r}{2}} -1)
({\rm xguess}^{\frac{r}{2}}+1).$  As long as ${\rm xguess}^{\frac{r}{2}}  $ is not one, 
at least one of  $({\rm xguess}^{\frac{r}{2}})^2 \pm 1$  must have a common factor 
with $M,$ and therefore finding ${\rm GCD }[ ({\rm xguess}^{\frac{r}{2}})^2 \pm 1]$  yields 
the factors of $M.$}
\be
f_1={\rm GCD }[{\rm xguess}^{r/2} +1,M] ; \qquad
f_2={\rm GCD }[{\rm xguess}^{r/2}-1,M]
\ee  
to determine the factors $f_1,f_2$ of $M.$  The above process simplifies 
if the ratio $\frac{2^{n_1}}{r} =D$ is already an integer.

In \QCMPI\  the local  probability maxima  and the associated factors 
are all stipulated.  In an actual measurement, one of those results would 
be found with that probability.

\section{PARALLEL UNIVERSE AND NOISE}
\label{sec7}

A real quantum computer will involve the manipulation of qubits using 
external fields and interactions with single  qubits and between qubits. 
Clearly, each physical realization has its set of Hamiltonians that 
describe that system and these QC manipulations. The circuit description of 
QC involves gates,  which in turn should be described by the action of 
Hamitonians on qubits.  For example, the simple one-qubit Hadamard gate 
can be realized by  rotating the  qubit's spin axis from the $\hat{z}$ to 
the  $\hat{x}$ axis  by means of a $-\vec{\mu}\cdot \vec{B}$ interaction 
acting for the proper time. More complicated gates involve clever design 
of one and two-qubit interactions.  In the future,  we hope that \QCMPI\ 
will provide a tool for describing all requisite gates based on Hamiltonian 
evolution. Dynamical evolution involves one- ( $H_1$) and two- ( $H_2 $) body 
Hamiltonians
$
\mid \Psi(t + \delta t) \rangle = [ 1 - \frac{i}{\hbar}  (  H_1 + H_2)  \delta t ]\  \mid \Psi(t ) \rangle
.$ 
Their action over a small time  interval $\delta t $  can be calculated 
by repeated application of the {\bf OneOpA} and {\bf TwoOpA} codes provided in \QCMPI. 
Such applications are the subject for future studies.

The major obstacle to the implementation of such gates required for the 
success of QC algorithms is the strong possibility that random intrusions, 
such as noise, will decohere the quantum system and remove the essential 
feature of quantum interference.  That issue behoves us to simulate the 
affect of noise by considering many replications of the QC algorithm, 
which ideally are identical,  and then subject each of them to random 
single and double one- qubit as well as single  two-qubit errors. For that 
task MPI is ideally suited and therefore,  as a major part of this paper, we 
have implemented that ``Parallel Universe" approach, for which we include
herein the Grover and Shor algorithms.  Other cases (teleportation and 
superdense coding) have also been implemented.  Subsequent numerical 
studies of the efficacy of error correction protocols can be implemented 
using the framework provided by the parallel universe feature of \QCMPI.

The next feature of this ``Parallel Universe" approach is that all of the 
state vector amplitudes can be gathered together and used to construct an 
ensemble average in the form of a density matrix. This process corresponds 
to solving a set of stochastic Schr\"{o}dinger equations~\cite{SSE} and using 
those solutions to produce a density matrix.  Let us now examine the steps 
needed to construct a density matrix.

\subsection{Density Matrix}
There are  advantages to using a density matrix to describe QC dynamics.  
The density matrix describes an ensemble average of quantum systems, with 
its evolution determined not only by the system's Hamiltonian but also by 
environmental terms using either  Kraus operator~\cite{Kraus} or 
Lindblad~\cite{Lindblad} differential equation forms  In addition,  the 
description of entanglement and of mixed states is handled nicely and 
concepts like entropy and Fidelity can be evaluated more readily. To form 
a density matrix in \QCMPI\ and to determine the entropy, affords a good 
example of how to extend  \QCMPI\  to such ensemble averages.
   
For a definite state vector, the pure state density matrix\footnote{The 
density matrix is Hermitian, has unit trace, and is positive definite.
In general $\rho^2 \leq \rho$,  with the equal sign applied for pure states.}
is simply
\be
 \rho = \mid  \Psi \rangle\langle \Psi \mid = \sum_{n=0}^{2^{n_q} -1} \  \   
\sum_{n'=0}^{2^{n_q} -1} C^*(n') C(n)  \mid n\rangle  \langle n' \mid
 \,  .
\ee
This large $(2^{n_q}\times 2^{n_q})$  matrix can be distributed over $N_P=2^p$  
processors by placing   $2^{n_q  -  p} \times 2^{n_q  -  p}$ matrices on each 
processor.~\footnote{To facilitate the parallel treatment of the density matrix, we take p as even.}
Matrix multiplication, traces and eigenvalue determination can 
then be implemented using MPI procedures, supplemented by BLACS processor grid
and parallel linear algebra SCALAPACK programs~\cite{scala}. Once the eigenvalues of $\rho$ 
are calculated the entropy can be determined.  But for a pure state,  we know 
that $\rho^2 \equiv \rho,$ and since the trace of $\rho$ is one, the
eigenvalues for a pure state are 1 and $2^{n_q}-1$ zeros. Thus the entropy is 
zero, as it should be for a well-defined, non-chaotic, albeit probabilistic state.
 
How do we go beyond a pure state density matrix within the \QCMPI\  setup?  
There are several options, but one overall goal. The overall goal is to 
build a state $\mid \Psi_\alpha \rangle$ repeatedly as labeled by $\alpha,$ 
with an associated probability ${\cal P}_\alpha$ with 
$\sum_\alpha\ {\cal P}_\alpha  =1.$ For each case, the state $\mid \Psi_\alpha
\rangle$ could be generated in a different way. One option is to get a set 
of amplitudes $C_\alpha (n)$ randomly,  with each random set assigned  a 
probability ${\cal P}_\alpha.$  Another way is to select a few qubits and
subject them to random one and two body interactions and possible stochastic 
pulses (noise),  again assigning each case   a probability ${\cal P}_\alpha.$  
The associated mixed state density matrix would then be 
\be
\rho =\sum_\alpha   {\cal P}_\alpha    \mid  \Psi_\alpha \rangle\langle\Psi_\alpha  \mid 
= \sum_{n=0}^{2^{n_q} -1} \ \  \sum_{n'=0}^{2^{n_q} -1}\ \    
\sum_\alpha   {\cal P}_\alpha\  C_{\alpha }^*(n') C_{\alpha }(n)  \mid
n\rangle  
\langle n' \mid \,  .
\ee
The above result can be expressed as~\footnote{ An  abbreviated version 
is  $\rho =\sum_\alpha \! {\cal P}_\alpha\!  \mid \alpha\rangle\langle\alpha\mid,$ 
with $C_{\alpha }(n) = \langle n \mid \alpha\rangle.$}
\be
\langle n \mid  \rho \mid n'\rangle =  \sum_\alpha   
{\cal P}_\alpha\  C_{\alpha }(n)\, C_{\alpha }^*(n')    \,  .
\ee 
This is  perhaps not the most general density matrix, but one can trace out 
some of the ancilla qubits and/or subject the density matrix  to additional 
entangling operations using $\rho'= U  \rho  U^\dagger$ or even apply the 
non-unitary  Lindblad~\cite{Lindblad}  process to generate an enhanced range 
of density matrices. These procedures, which we outline here, are included 
in this version of \QCMPI \  to facilitate studies of decoherence and
environmental effects. A major advantage of \QCMPI\ is that the invocation of
 parallel universes (aka multiverses) to describe the influence of noise on 
a QC does not involve much increase in computation time compared to a single 
pure run, especially  since the only communication  between groups is that 
used is to construct the density matrix.
 This scheme provides an efficient use of multi-processor computers.
 
 \subsection{Parallel universe implementation}
 
The above steps are implemented in \QCMPI \ by first splitting 
the overall number of processors $N_P$(nprocU) into many groups 
$N_G$,  each group is referred to as a ``multiverse."  For 
convenience, we take both $N_P$ and $N_G$ to be powers of 2. 
Within each multiverse, there are $N_P/N_G \equiv N_g$ (nprocM) 
processors that are used to perform a distinct QC algorithm. The 
MPI command {\bf MPI\_COMM\_SPLIT}  is used to produce these 
separate groups. Each group is specified by its group rank (rankM), 
which ranges from zero to NGROUPS-1, where NGROUPS denoted the 
total number of multiverses. ~\footnote{There are spawning features 
of MPI-2 that might be invoked to carry out this process more 
efficiently, but at this stage we found MPI-1 sufficient for our needs.}
 
The method used to store and evaluate the density matrix is controlled by 
an integer {\bf  Ientropy}. For the choice {\bf Ientropy}=0, there 
is no evaluation of the density matrix. For the choice {\bf Ientropy}=1, 
the full density matrix is constructed on the master processor and its 
eigenvalues determined by a LAPACK code. That procedure should be used 
when storage space for $\rho$ is ample. For {\bf Ientropy}=2 the density matrix 
is not stored on one processor, but is distributed on a BLACS generated 
processor grid and  the parallel eigenvalue code {\bf PCHEEVX} from 
the SCALAPACK package is invoked to evaluate $\rho 's$ eigenvalues. To 
carry out this last task the number of processors, groups and qubits 
have to be carefully monitored for consistency with the codes 
conventions, as indicated directly in the listings.

  \subsection{Noise scenarios}
  
A simple example of a ``noise scenario"  has been included ~\footnote{See 
subroutine Noise called in subgrover.f90 and subshor.f90}in  \QCMPI \  to 
show how the role of noise can be examined. The motivation here is to first 
introduce noise and later to evaluate various error correction  schemes.
  
The division of a large number of processors into groups was made so that 
only the first group (rankM=0) functions without noise.  All of the other 
groups perform the algorithm with noise.  The noise is introduced separately 
for each group (or multiverse) where the users can design their own scenarios.
We have input noise using a one-qubit  unitary operator (subroutine D2) that 
we take as a $2 \times 2$  Wigner rotation function 
${\cal D}^{\frac{1}{2}}(\alpha,\beta,\gamma),$ where $\alpha,\beta,\gamma$ are 
three Euler angles.~\footnote{  We take random  $\alpha,\beta$, and set 
$\gamma=0$  for simplicity.} This can be specialized to either small 
deviations or, within a phase, to one of the Pauli operators. One can introduce 
one qubit  noise, acting on a random qubit, typically once within each 
processor multiverse, but two or more one-qubit noise intrusions can be 
invoked at various stages of the algorithm,  by suitable placement of the subroutine ``Noise."
  
In addition, a two-qubit  unitary operator (subroutine D4) that we take 
as a $4 \times 4$  Wigner function ${\cal
  D}^{\frac{3}{2}}(\alpha,\beta,\gamma),$ can also be specialized to either 
small deviations or,  within a phase, to one of the Pauli operator products 
$\sigma_i \otimes \sigma_j .$  This allows one to introduce a single error 
that acts on two qubits once, in contrast to two one-qubit errors.
  
The one-qubit operator is assumed to act on a random selected qubit (qhit) 
and at selected, variable stages of the algorithm (eloc).  Extension to 
two-qubit noise is obvious. Of course the associated universes which allow 
two one-qubit or single two-qubit errors should carry lower weight.
  
By using unitary operators in each universe the overall density matrix 
still maintains unit trace,  but of course the trace of $\rho^2$ will be 
decreased by noise. The probability of success will also decrease.
  
Thus, \QCMPI \   provides a framework for introducing errors and, along 
with Hamiltonian-driven gates, provides an important tool for dynamical 
studies of QC with noise and  in the future with error correction.

\section{FORTRAN AND  MPI CODES}
\label{sec8}

 Sample \QCMPI \  codes are provided which incorporate directions as to how to
run the code.  From these examples, the user should be able to see the benefit 
of being able to handle problems with a considerable number of qubits, 
organized into parallel universes, in reasonable time. Some improvements could 
be invoked to accelerate \QCMPI \ , for example  by collecting messages and
sending them as a group (collective communications).  The issue here is the 
standard fight between sharing the work load over the available processors 
(balance)  and minimizing  the cost of sending messages.  However, the major benefit of dividing a large number of processors
into multiverses and subjecting each one to separate noise scenarios, is 
in itself justification and reason to use a multi-processor supercomputer.

The list of files contained in QCMPI are:
\begin{itemize}
\item qcmpisubs.f90, contains all QCMPI subroutines  
\item Guniv.f90, builds multiverse environment for Grover's search
\item subgrover.f90, Grover's search routine
\item Suniv.f90, builds multiverse setup for Shor's factoring algorithm
\item subshor.f90, Shor's factoring routine
\item makefile, sample of compiling options for several 
supercomputing facilities
\item *.job, sample job submitting scripts
\item README, instructions.
\end{itemize}

\subsection{Performance}
As an indication of the performance of  \QCMPI,  we have run a number 
of  sample cases using the multiverse Grover codes included in the package. In 
table~\ref{tablep},  we show the global memory requirements together 
with the wallclock time and the percentage of the latter used in MPI 
operations. 

\begin{table}[h]
\caption{Performance of a number of sample runs  all realized using 
Guniv.f, subgrover.f90 and qcmpisubs.f90. In all cases.  the 
Ientropy=2 option is chosen and thus the entropy is computed making 
use of the scalapack routines and two multiverses are considered. 
The Gflop/sec and Gbytes refer to the total amount used by all processors.
The information presented has been obtained using IPM~\cite{ipm}.
\label{tablep}}
\vspace{10pt}
\begin{tabular}{l | llllrr}
\hline
NP   &  $nq$   & $2^{nq}$  & Gflop/sec  
 & Gbytes  & Wallclock (sec)   &  \% communication  \\
\hline
\hline
4          &  10  & 1024  & 1.99618 & 1.4043  & 1.25 & 26.63  \\
4          &  12  & 4096  & 3.31855 & 9.96069 & 48.98 & 5.65  \\
16         &  10  & 1024  & 1.90923 & 5.54911 & 1.07 & 62.27    \\
16         &  12  & 4096  & 10.4583 & 38.7235 & 11.29 & 39.14   \\
64         &  10  & 1024  & 0.65133 & 22.1393 & 3.21 & 31.19    \\
64         &  12  & 4096  & 15.2444 & 153.814 & 7.54 & 56.32   \\
\hline
\hline
\end{tabular}
\end{table}

\newpage

\section{CONCLUSIONS AND FUTURE DEVELOPMENTS}
\label{sec9}
   
In conclusion,  the Fortran 90 code \QCMPI \  provides a modular approach to
quantum algorithms that provides accessible implementation of quantum computation
algorithms.  All of the gates needed for the circuit model are provided, as 
well as the quantum Fourier transformation procedure.  Extension to
three-qubit 
operators and to the  one-way model of computation are
straightfoward,  as is the extension to the qutrit case.  Such extensions will
be provided in the future by the
authors and hopefully also by interested users. 

The main features of \QCMPI \ are the distribution of state-vector amplitudes
over processors, to allow for increased number of qubits and the use of MPI to 
carry out the requisite communication needed when one- and two- body operators 
(gates) act on states.  This task is carried out in a manner that allows 
ready extension to Hamiltonian driven QC dynamics.

In addition.  \QCMPI \  provides a multi-universe setup,  which replicates 
the QC algorithm over many groups, at little cost in computation time. That  
procedure provides a major advantage of \QCMPI\, , not generally available in 
the literature, to provide a framework for studying the role of noise on the 
efficacy of QC.  That is,  we believe,  the major task in this subject.

The methods demonstrated here for the distribution and evaluation of a large density matrix
can be generalized to the case of large unitary matrices to represent gates.

There is much to do with this tool such as studies of:  Hamiltonian driven QC dynamics
using realistic Hamiltonians,  along with environmental effects,  influence of
random pulses,  and efficacy of error correction protocols.

\section*{Acknowledgments}
We gratefully acknowledge the help and participation of Prof. C.W. Finley at 
an early stage of this work. This project was supported  in part by the U.S. National
Science Foundation and in part under  Grants  PHY070002P  \&  PHY070018N  from
the Pittsburgh Supercomputing Center, which is supported by several federal
agencies, the Commonwealth of Pennsylvania and private industry. Circuit
graphs were prepared using codes from Ref.~\cite{drawings} which we
appreciate. Thanks to PSC staff members Dr. Roberto Gomez and  Rick Costa. We 
also thank the openmpi and scalapack groups, especially Julie Langou  and  Jeff Squyres   .  
 
\newpage
\appendix

\section{The quantum Fourier transform circuit and \QCMPI}

\begin{figure}[hb]
\begin{center}
\includegraphics[width=14cm]{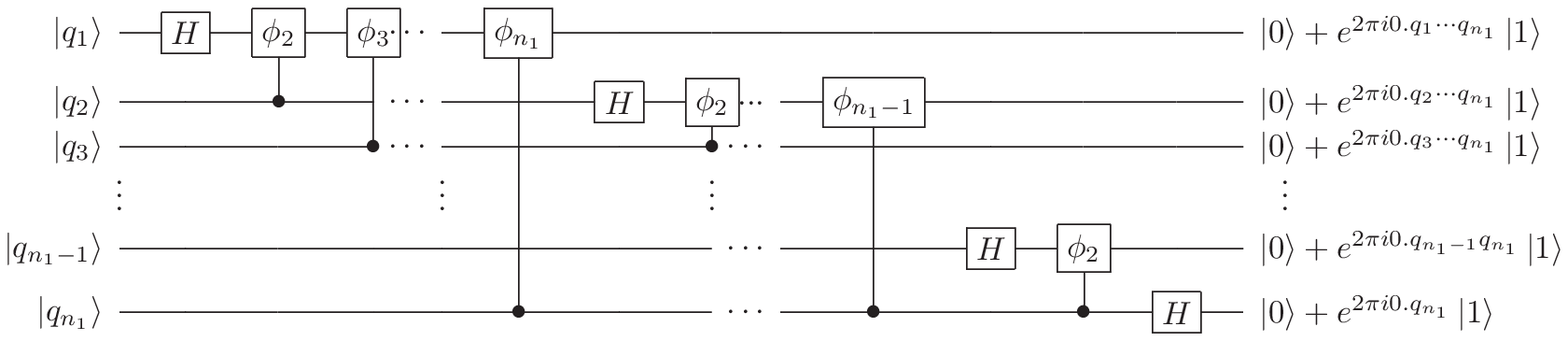}
\caption{The quantum Fourier transform circuit. 
Here $ \phi_k\equiv e^{ 2 \pi  i /2^k}$ and the register 
one has $n_1$ qubits.The binary number $q_1 q_2 \cdots q_{n_1}$ 
corresponds to a decimal number $n$ which ranges as $0\leq n\leq 2^{n_1} -1.$ 
The fraction binary notation is used where  
$0.q_a\cdots q_{b} \equiv \frac{q_a}{2} + \frac{q_{a+1}}{2^2} +\cdots
+\frac{q_b}{2^{  b-a+1}}.$  To restore the qubits to standard order with the
most significant bit to the left (top of figure),  an addition reversal of the
qubit order must be applied. An overall normalization of $1/2^{\frac{n_1}{2}}$  is understood. 
}
\end{center}
\label{QFTcircuit}
\end{figure}

The quantum Fourier transform is performed using the circuit in
Fig.~\ref{QFTcircuit}. A ladder of Hadamards and two-qubit control 
CPHASEK gates (Eq.~(\ref{cphasek})) are used to produce the QFT.

The above steps are carried out in \QCMPI\ with the following code, which
includes a series of pair swaps  to reorder the qubits labels. 

\begin{verbatim}
Do ic =1,n1-1
call OneOpA(nq,ic,had,psi,NPART,COMM)
Do k=ic+1,n1
call CPHASEK(nq,k,ic,k+1-ic,psi,NPART,COMM)
enddo
enddo
! Final Hadamard
call OneOpA(nq,n1,had,psi,NPART,COMM)

! Reverse order using pair swaps
Do i=1,n1/2
call SWAP(nq,i,n1+1-i,Psi,NPART,COMM)
enddo
\end{verbatim} 

Here {\bf had} denotes the Hadamard, {\bf psi} is the input and
then the output state vector at each stage, and {\bf NPART} denotes the 
part of {\bf psi} on the current processor. The qubits are restored 
to standard order by a set of pair swaps. This also demonstrates how 
to use the  {\bf CPHASEK, OneOP}(for a Hadamard case),and the {\bf SWAP} subroutine:

To understand how the ladder of Hadamard and control phase  gates  yields a
quantum Fourier transform,  note that the final state shown in the code, 
\bea
\noindent \mid q_1  q_2 \cdots q_{n_1}  \rangle   \rightarrow \frac{1}{\sqrt{2^{n_1}}}   (  \mid 0\rangle + 
 e^{2 \pi i \ 0.q_1 \cdots q_{n_1} }  \mid 1  \rangle ) \otimes  
  (  \mid 0\rangle +  e^{2 \pi i \ 0.q_2 \cdots q_{n_1} }  \mid 1  \rangle )  &&      \nonumber \\
\otimes \cdots (  \mid 0\rangle +  e^{2 \pi i \ 0.q_{n_1-1}  q_{n_1} }  \mid 1  \rangle ) \otimes
 (  \mid 0\rangle +  e^{2 \pi i \ 0.q_{n_1}  }  \mid 1  \rangle )&& ,  \nonumber \\
 && 
\eea 
which becomes
\bea
\noindent \mid q_1  q_2 \cdots q_{n_1}  \rangle   \rightarrow   \frac{1}{\sqrt{2^{n_1}}} (  \mid 0\rangle + 
 e^{2 \pi i \ 0. q_{n_1} }  \mid 1  \rangle ) \otimes  
  (  \mid 0\rangle +  e^{2 \pi i \ 0.q_{n_1-1}   q_{n_1} }  \mid 1  \rangle )  &&      \nonumber \\
\otimes \cdots (  \mid 0\rangle +  e^{2 \pi i \ 0.q_2  \cdots q_{n_1}}  \mid 1  \rangle ) \otimes
 (  \mid 0\rangle +  e^{2 \pi i \ 0.q_1 \cdots q_{n_1}  }  \mid 1  \rangle )&& ,  \nonumber \\
 && 
\eea    
after the final qubits  are reordered by a series of pair swap operations.  The last result can be written as:
\be
  {\rm QFT} \mid q_1  q_2 \cdots q_{n_1}  \rangle=  \frac{1}{\sqrt{2^{n_1}}}
  \sum_{Q'}   e^{2 \pi i \   [ q'_1 0. q_{n_1} +  q'_2 0. q_{n_1-1}   q_{n_1}   \cdots   
 + q'_{n_1-1} 0.q_2  \cdots q_{n_1}  + q'_{n_1} 0.q_1  \cdots q_{n_1} ]  }
 \    \mid Q'  \rangle,
\ee  
where $ Q'$ denotes the binary number $ q'_1  q'_2 \cdots q'_{n_1},$
corresponding to the decimal number $n'$.  The above is equivalent to
\be
{\rm QFT} \mid n \rangle=  \frac{1}{\sqrt{N}}  \sum_{n'=0}^{N-1}   e^{ 2 \pi i n\, n'/N}   \mid n' \rangle.
\ee  
where $N= 2^{n_1} .$  A simple product $ n\, n'/N$ appears in the exponent
because 
$ [ q'_1 0. q_{n_1} +  q'_2 0. q_{n_1-1}   q_{n_1}   \cdots  + q'_{n_1-1}
  0.q_2  \cdots q_{n_1}  + q'_{n_1} 0.q_1  \cdots q_{n_1} ] \rightarrow n
n'/N,$  which can be shown by noting that $ e^{ 2 \pi i 2^s} \equiv 1$ for all
integers $s \geq 0.$ Therefore, in the product $n n'/N = ( q_1 2^{n_1-1} + q_2 2^{n_1-2} \cdots q_{n_1} 2^0)( q'_1 2^{n_1-1} 
+ q'_2 2^{n_1-2} \cdots q'_{n_1} 2^0),$ we can drop all cross terms that yield
a $2^s$ which suffices to prove the equivalence. Hence, one sees that the 
{\rm QFT} is a unitary transformation from basis $\mid n\rangle$ to $\mid
n'\rangle$ of the form $\langle n \mid  {\rm QFT}  \mid n'\rangle =   
\frac{1}{\sqrt{N}}\  (e^{ 2 \pi i  /N} )^{n n'}. $

\section{The MPI Codes}

The 
{\bf bintodec,
dectobin,
OneOpA,
TwoOpA,
EulerPhi,
splitn,
ProjA,
Randx,
QFT,
CF,
SWAP,
CPHASEK,
HALL,
HALL2,
Entropy,
EntropyP} codes are best understood by examination of
the explicit directions within the code and also by the useage in the sample algorithms.

\subsection{ Sample algorithm codes} 
Teleportation and Superdense coding codes are also available.  In this 
paper, we present the Grover and Shor cases.   The Grover code is called
Guniv.f90 and initiates the process by selecting a marked item that is to be
searched for, with that item (labelled as IR) distributed to all the
processors. There are $N_P=2^p$ processors that are split into $N_G= 2^g$ groups
(called multiverses) each multiverse then consists of $N_x=N_P/N_G=2^{p-g}$ members, where 
both $N_P$ and $N_G$ are assumed to be powers of 2. Independent searches are
carried out in each multiverse and at the end  (this could be done at any
preferred stage) the state-vector amplitudes for each group are used to form a 
group's density matrix.  An overall ensemble average of all the group's
density matrices are then computed and either  the $2^{n_q}\times 2^{n_q}$
array  is located on the master processor of the first group using subroutine 
{\bf Entropy} or
is distributed over a BLASC grid using subroutine {\bf EntropyP}.

The first group (rankM$=$0) is free of noise, whereas all the other groups 
(rankM$>$0) are subject to various random disturbances with assigned 
probablities. This is where particular noise models could be invoked by 
the user. This structure is also used for the Shor case.
  
In the Shor case (Suniv.f90, subshor.f90),  there is an initial setup process to pick and test 
the number to be factored that is broadcast  to all $N_P$ processors,  with
again a split into $N_G$ groups (multiverses) and separate searches done on the 
$N_g$ members of each universe. Again group one is free of noise, whereas
noise is introduced on all other groups,  with a subsequent build up of the 
full density matrix using either the {\bf ientropy=1} or   {\bf ientropy=2} options. Others cases and extensions all follow this same general pattern.
  
One can also examine the particular eigenvalues of the full density matrix,
at selected stages, and also obtain fidelities and subtraces if desired.

\end{document}